\documentclass[twocolumn,superscriptaddress,secnumarabic,
amssymb,amsmath,nobibnotes,aps,prd,showpacs,nofootinbib]{revtex4}%
\usepackage{graphicx}
\usepackage{epsf}
\usepackage{bm}
\usepackage{amsmath}
\usepackage{amsfonts}
\usepackage{amssymb}
\usepackage{epstopdf}
\usepackage{natbib}%
\usepackage{rotating}
\usepackage{pdflscape}
\usepackage{adjustbox}
\providecommand{\U}[1]{\protect\rule{.1in}{.1in}}
\newcommand{\be}{\begin{equation}}
\newcommand{\ee}{\end{equation}}

\newcommand{\mincir}{\raise
-3.truept\hbox{\rlap{\hbox{$\sim$}}\raise4.truept\hbox{$<$}\ }}
\newcommand{\magcir}{\raise
-3.truept\hbox{\rlap{\hbox{$\sim$}}\raise4.truept\hbox{$>$}\ }}

\usepackage{color}

\begin{document}

\title{Metastable dark energy models in light of Planck 2018: Alleviating the $H_0$ tension}

\author{Weiqiang Yang}
\email{d11102004@163.com}
\affiliation{Department of Physics, Liaoning Normal University, Dalian, 116029, P. R. China}

\author{Eleonora Di Valentino}
\email{eleonora.divalentino@manchester.ac.uk}
\affiliation{Jodrell Bank Center for Astrophysics, School of Physics and Astronomy,
University of Manchester, Oxford Road, Manchester, M13 9PL, UK.}

\author{Supriya Pan}
\email{supriya.maths@presiuniv.ac.in}
\affiliation{Department of Mathematics, Presidency University, 86/1 College Street,
Kolkata 700073, India.}

\author{Spyros Basilakos}
\email{svasil@academyofathens.gr}
\affiliation{Academy of Athens, Research Center for Astronomy and Applied Mathematics, Soranou Efessiou 4, 115 27 Athens, Greece. }
\affiliation{National Observatory of Athens, Lofos Nymfon,
11852, Athens, Greece. }

\author{Andronikos Paliathanasis}
\email{anpaliat@phys.uoa.gr}
\affiliation{Institute of Systems Science, Durban University of Technology, PO Box 1334, Durban 4000, Republic of South Africa}
\affiliation{Instituto de Ciencias F\'{\i}sicas y Matem\'{a}ticas, Universidad Austral de Chile, Valdivia 5090000, Chile}



\begin{abstract}
We investigate the recently introduced metastable dark energy (DE) models after the final Planck 2018 legacy release. The essence of the present work is to analyze their evolution at the level of perturbations.
Our analyses show that both the metastable dark energy models considered in this article,  are excellent candidates to alleviate the $H_0$ tension. In particular, for the present models, Planck 2018 alone can alleviate
the $H_0$ tension within 68\% CL.  Along with the final cosmic microwave background data from the Planck 2018 legacy release, we also include external cosmological datasets in order to asses the robustness of our findings.
\end{abstract}

\maketitle
\section{Introduction}

The nature of dark energy (DE) or geometrical dark energy (GDE) is one of the intrinsic queries of modern cosmology that we are still looking for. According
to the analyses of the high quality observational data, the present accelerating phase of the universe is quite well described in the framework of the general relativity together with a cosmological constant -- the so called $\Lambda$CDM model. However, due to many theoretical and observational shortcomings associated with the $\Lambda$CDM cosmology, searches for alternative descriptions have been necessary. Apart from the well known cosmological constant/fine tuning and cosmic coincidence problems affecting the $\Lambda$CDM scenario, recent observations indicate that the CMB measurements of some key cosmological
parameters within this minimal $\Lambda$CDM scenario do not match with the values measured by other cosmological probes.
Specifically, one is the long standing $H_0$ tension (above 4$\sigma$) between the estimated value of $H_0$ provided by Planck \cite{Aghanim:2018eyx} (in agreement with with~\cite{Gott:2000mv,Chen:2011ab,Efstathiou:2013via,Chen:2016uno,Wang:2017yfu,Lin:2017bhs,Yu:2017iju,Abbott:2017smn,Haridasu:2018gqm,Zhang,Park:2018tgj,Zhang:2018air,Ryan:2019uor,Dominguez:2019jqc,Cuceu:2019for,Lukovic:2019ryg,Zeng:2019mae,Lin:2019zdn,Freedman:2019jwv,Freedman:2020dne,Cao:2020jgu,Alam:2020sor,Birrer:2020tax}) and that one measured by the SH0ES collaboration~\cite{Riess:2019cxk} (see also~\cite{Rigault:2014kaa,Cardona:2016ems,Riess:2016jrr,Zhang:2017aqn,Dhawan:2017ywl,Fernandez-Arenas:2017isq,Birrer:2018vtm,Riess:2018uxu,Camarena:2019moy,Wong:2019kwg,Yuan:2019npk,Huang:2019yhh,Shajib:2019toy,Verde:2019ivm,Henning:2017nuy,Reid:2019tiq,Pesce:2020xfe}.
The other one is the $S_8$ tension between Planck and  the cosmic shear measurements KiDS-450~\cite{Kuijken:2015vca,Hildebrandt:2016iqg,Conti:2016gav}, Dark Energy Survey (DES)~\cite{Abbott:2017wau,Troxel:2017xyo} or CFHTLenS~\cite{Heymans:2012gg, Erben:2012zw,Joudaki:2016mvz}. Furthermore, when a curvature is considered into the cosmic picture~\cite{DiValentino:2019qzk}, all these tensions are exacerbated revealing a possible crisis for the cosmology. Thus, in order to circumvent
these problems, several alternative cosmological models have been introduced in the literature aiming to solve or alleviate such tensions in an effective way. In the literature there is a large family of models that alleviate
the $H_0$ tension among which ``multi-parameter'' dark energy~\cite{DiValentino:2015ola,DiValentino:2016hlg,DiValentino:2017zyq,DiValentino:2019dzu,DiValentino:2020hov}, early dark energy~\cite{Pettorino:2013ia,Poulin:2018cxd,Alexander:2019rsc,Sakstein:2019fmf,Niedermann:2019olb,Ye:2020btb}, interacting dark energy~\cite{Kumar:2016zpg,DiValentino:2017iww,Kumar:2017dnp,Yang:2018euj,Yang:2018uae,Yang:2019uzo,Martinelli:2019dau,DiValentino:2019ffd,DiValentino:2019jae,DiValentino:2020evt}, modified gravity models~\cite{Raveri:2019mxg,Yan:2019gbw,Frusciante:2019puu}, and the list goes on (see~\cite{DiValentino:2017oaw,DiValentino:2017rcr,Khosravi:2017hfi,Renk:2017rzu,DiValentino:2017gzb,Fernandez-Arenas:2017isq,Sola:2017znb,Nunes:2018xbm,Colgain:2018wgk,DEramo:2018vss,Guo:2018ans,Yang:2018qmz,Banihashemi:2018oxo,Banihashemi:2018has,Zhang:2018air,Kreisch:2019yzn,Vattis:2019efj,Kumar:2019wfs,Agrawal:2019lmo,Yang:2019jwn,Yang:2019qza,DiValentino:2019exe,Desmond:2019ygn,Yang:2019nhz,Pan:2019gop,Visinelli:2019qqu,Martinelli:2019krf,Cai:2019bdh,Pan:2019hac,Schoneberg:2019wmt,Shafieloo:2016bpk,Li:2019san,Cuceu:2019for,Colgain:2019joh,Pan:2019jqh,Berghaus:2019cls,DiValentino:2020naf}. On the other hand, for the well known $S_8 = \sigma_8 \sqrt{\Omega_{m0}/0.3}$ tension we refer the reader the following works \cite{Pourtsidou:2016ico,An:2017crg,Kumar:2019wfs,DiValentino:2018gcu,Kazantzidis:2018rnb,DiValentino:2019ffd,DiValentino:2019dzu}. The above family of models provide a framework of alleviating such tensions within 3$\sigma$, but the problem still remains open.

In this article we consider two metastable DE models introduced recently by Shafieloo et al. \cite{Shafieloo:2016bpk} (also see \cite{Li:2019san}). The basic ingredient of these models is that the decay of DE does not depend on the external parameters, such as the expansion rate of the universe etc. These models depend only on the intrinsic properties of DE. Thus, it is expected that metastable DE models could explore some inherent nature
of the dark sector, specially the DE. Our observational constraints on the metastable DE models
should be considered stringent for the following reasons: (i) we have considered the cosmological perturbations for the models, an indispensable tool to understand the large scale structure of the universe, and (ii) we have included the final Planck 2018 data
\cite{Aghanim:2018eyx,Aghanim:2018oex,Aghanim:2019ame}. A quick observation from our analyses is that the metastable DE models are able to alleviate the $H_0$ tension.

The article is organized in the following way. In section \ref{sec-2}, assuming the Friedmann-Lema\^{i}tre-Robertson-Walker (FLRW) universe, we present the gravitational equations and two  metastable DE models that we wish to study in this work.
In section \ref{sec-data} we discuss the observational data and the methodology applied to constrain the models. Then we discuss the results of our analyses in section \ref{sec-results}. Finally, in section \ref{sec-discuss} we close our work with a brief summary of all the findings.

\section{Metastable dark energy models}
\label{sec-2}

In this section we review two metastable DE models introduced recently by \cite{Shafieloo:2016bpk,Li:2019san}.
We assume the spatially flat Friedmann-Lema\^{i}tre-Robertson-Walker (FLRW) geometry which is characterized by the line element
$ds^2 = - dt^2 + a^2 (t) \left[dx^2 + dy^2 +dz^2\right]$, where $a(t)$ (hereafter $a$) is the scale factor of the universe.
The gravitational sector of the universe follows Einstein's General Relativity where in addition we assume that the matter content of the
universe is minimally coupled to gravity. Further, we assume that the entire universe is comprised of baryons, radiation, pressure-less dark matter
and a dark energy fluid. Throughout the present work we shall identify $\rho_i$ and $p_{i}$ as the energy density and pressure of the $i$-th fluid.
Here, $i  = \{b, r, c, x \}$ stands for baryons ($b$), radiation ($r$), pressure-less or cold dark matter ($c$) and DE ($x$).
Within this framework, one could write down the Einstein's field equations:

\begin{eqnarray}
3 H^2  = \frac{8\pi G}{3} \sum_{i} \rho_i,\label{EFE1}\\
2\dot{H} + 3 H^2  = -4 \pi G \sum_{i} p_i,\label{EFE2}
\end{eqnarray}
where an overhead dot denotes the derivative with respect to the cosmic time; $H \equiv \dot{a}/a$ is the Hubble rate of the FLRW universe and $8 \pi G$ is the Einstein's gravitational constant ($G$ is the Newton's gravitational constant).  Let us note that using either the Bianchi's identity or using the gravitational equations (\ref{EFE1}) and (\ref{EFE2}), one could derive the conservation equation of the total fluid

\begin{eqnarray}\label{cons-total}
\sum_{i} \dot{\rho}_i  + 3 H \sum_{i} (\rho_i + p_i)  = 0~.
\end{eqnarray}
So, out of the three equations, namely, eqns. (\ref{EFE1}), (\ref{EFE2}) and (\ref{cons-total}), only two of them are independent. Since DE plays a crucial role in the dynamics of the universe, over the last two decades, several forms of DE have been studied in the literature. In most of the cases, it has been assumed that DE density depends on the external parameters, such as the  scale factor, $a$, of the FLRW universe; its expansion rate, $H$; or its scalar curvature. While one may naturally consider a scenario in which DE depends from its intrinsic composition and
structure. The motivation of the metastable DE models is along the latter lines. In the following we shall introduce two metastable DE models and discuss their physical origin.
\bigskip

\subsection{Model I}

The first metastable DE model that we aim to study follows the evolution law \cite{Shafieloo:2016bpk,Li:2019san}:

\begin{eqnarray}\label{model1}
\dot{\rho}_x = - \Gamma \rho_x~,
\end{eqnarray}
where $\rho_x$, as already mentioned, denotes the energy density of DE and $\Gamma$ is a constant which could be either positive or negative and its dimension is same as that of the Hubble rate, $H$, of the FLRW universe.
Note that, $\Gamma = 0$ implies $\rho_x =  \mbox{constant}$, featuring the cosmological constant. Note further that other cosmic fluids, namely baryons, radiation and cold dark matter follow the usual conservation equation, that means, $\dot{\rho}_i + 3 H (p_i +\rho_i) = 0$, where $i = \{b, r, c\}$.
The evolution of DE characterized in  eqn.~(\ref{model1}) is exponential, and for $\Gamma> 0$ DE density has a decaying character, while for $\Gamma < 0$ DE density is increasing. This kind of evolution is actually motivated from the `radioactive decay' scheme in which unstable nuclei and elementary particles may decay. Moreover, as we have already mentioned,
the energy densities of radiation, baryons, and cold dark matter obey the standard scaling laws implying that this model
can be viewed in the context of dynamical
dark energy. Hence, one can introduce a homogeneous scalar field $\phi$~\cite{Peebles:1987ek,Ratra:1987rm} rolling down the potential energy $V(\phi)$, and therefore it could
resemble a scalar field model of DE.
Now, if we focus on the evolution of DE as given in eqn. (\ref{model1}), that means, $\dot{\rho}_x + \Gamma \rho_x = 0$, one could quickly find its equivalent structure by comparing it with the standard evolution of DE
\begin{eqnarray}\label{DE-normal}
\dot{\rho}_x + 3 H (1+w_x) \rho_x = 0,
\end{eqnarray}
which naturally introduces a dynamical equation of state of DE, $w_x = p_x/\rho_x$. Thus, comparing (\ref{model1}) and (\ref{DE-normal}), one
could determine, $w_x  = -1 + \frac{\Gamma/H_0}{3H/H_0}$, where we introduce $H_0$, i.e. the present value of $H$. In other words, $\Gamma$ will give us an estimate of the deviation of the dark energy equation of state from the cosmological constant.

Let us now proceed with the evolution of this model at the level of perturbations. Here
we consider the perturbed FLRW metric  in the synchronous gauge \cite{Ma:1995ey}
\begin{eqnarray}
\label{perturbed-metric}
ds^2 = a^2(\tau) \left [-d\tau^2 + (\delta_{ij}+h_{ij}) dx^idx^j  \right],
\end{eqnarray}
where $\tau$ is the conformal time; $\delta_{ij}$,  $h_{ij}$
respectively denote the unperturbed and perturbed metric  tensors. Now, for the above metric (\ref{perturbed-metric}), using the conservation equation for the total fluid, one can conveniently derive the corresponding evolution equations Fourier space $k$, and they are

\begin{widetext}
\begin{eqnarray}
&&\delta_x^{\prime} = - (1+w_x) \left(\theta _x + \frac{h^{\prime}}{2}\right) - 3 \mathcal{H} (c_{sx}^2 - w_x) \left[\delta_x + 3 \mathcal{H} (1+w_x) \frac{\theta_x}{k^2}\right] - 3 \mathcal{H} w_x^{\prime} \frac{\theta_x}{k^2}~, \label{pert1}\\
&&\theta_x^{\prime} =  - \mathcal{H} (1- 3 c_{sx}^2) \theta_x + \frac{c_{sx}^2}{1+w_x} k^2 \delta_x~, \label{pert2}\\
&&\delta_c^{\prime} = - \left(\theta _c + \frac{h^{\prime}}{2}\right)~,\label{pert3}\\
&&\theta_c^{\prime} =  - \mathcal{H} \theta_c~,\label{pert4}
\end{eqnarray}
\end{widetext}
where the primes attached to any quantity denote the derivative of that quantity with respect to the conformal time $\tau$; $\mathcal{H}= a^{\prime}/a$, denotes the conformal
Hubble factor; $h = h^{j}_{j}$ is the trace of the metric perturbations $h_{ij}$; $\theta_{i}\equiv i \kappa^{j} v_{j}$ (here $i =  c, x$) is the divergence of the $i$-th fluid
velocity.  Finally, $\delta_i = \delta \rho_i/\rho_i$ denotes the density perturbation for the $i$-th fluid, that means $\delta_x$ is the density perturbation for the dark energy fluid while $\delta_c$ refers to the density perturbation for the cold dark matter fluid.  Notice that $c^2_{sx}=\delta p_x/\delta\rho_x$, is the effective sound speed of the DE perturbations in the rest frame \cite{Hu:1998kj}  (the corresponding quantity for
matter is zero in the dust case), which determines the amount of DE clustering  and it can be treated as a free parameter without any problem.
However, we need to have in mind that the inclusion of the sound speed as a free parameter actually increases the degeneracy among the model parameters.
On the other hand, for barotropic DE with constant equation of state $w_x$, $c^2_{sx} = w_x <0$, and hence instabilities appear in the DE fluid
\cite{Valiviita:2008iv,Yang:2017zjs}.  In order to avoid instabilities one has to impose $c_{sx}^2 > 0$ \cite{Valiviita:2008iv,Yang:2017zjs}.
It is well known that in the case of a homogeneous dark energy we have $c_{sx}^2 =1$, hence, the corresponding pressure suppresses any DE fluctuations at sub-horizon scales, and consequently,  the quantities $\delta_x$  and $\theta_{x}$ are   vanished. On the other hand, for $c_{sx}^2 =0$, DE clusters similar to that of dark matter perturbations. The clustering of DE modifies
the evolution of dark matter fluctuations perturbations
(for more discussion see \cite{Ballesteros:2008qk,Sapone:2012nh,Pace:2014taa,Basilakos:2014yda,Nesseris:2015fqa,Mehrabi:2015hva} and the references therein).  In the current paper we have set $c_{sx}^2 =1$,
which implies that dark energy is non-clustering, hence one
should consider the perturbation equations  along with
the background ones.

\begin{figure*}
\includegraphics[width=0.45\textwidth]{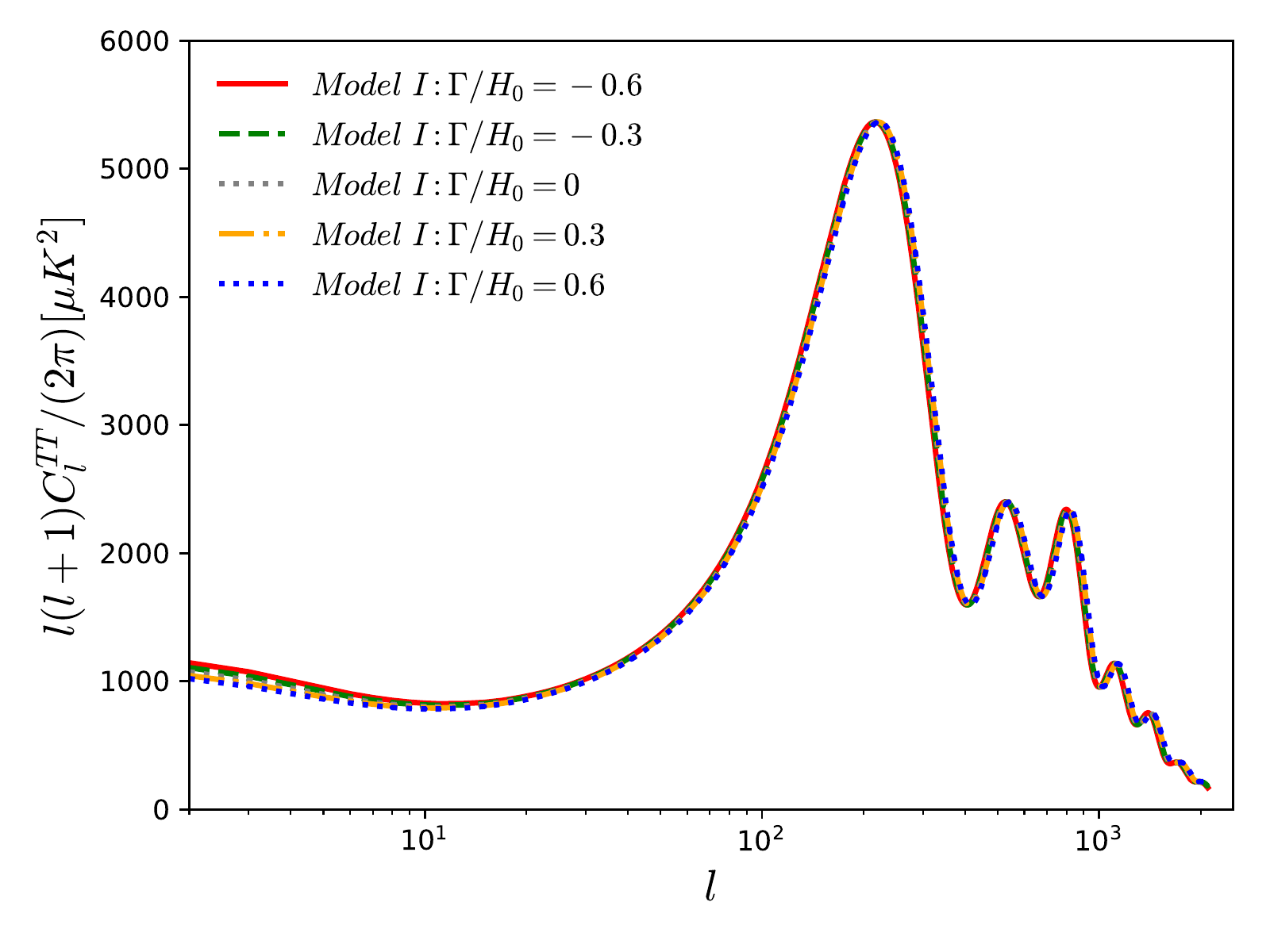}
\includegraphics[width=0.45\textwidth]{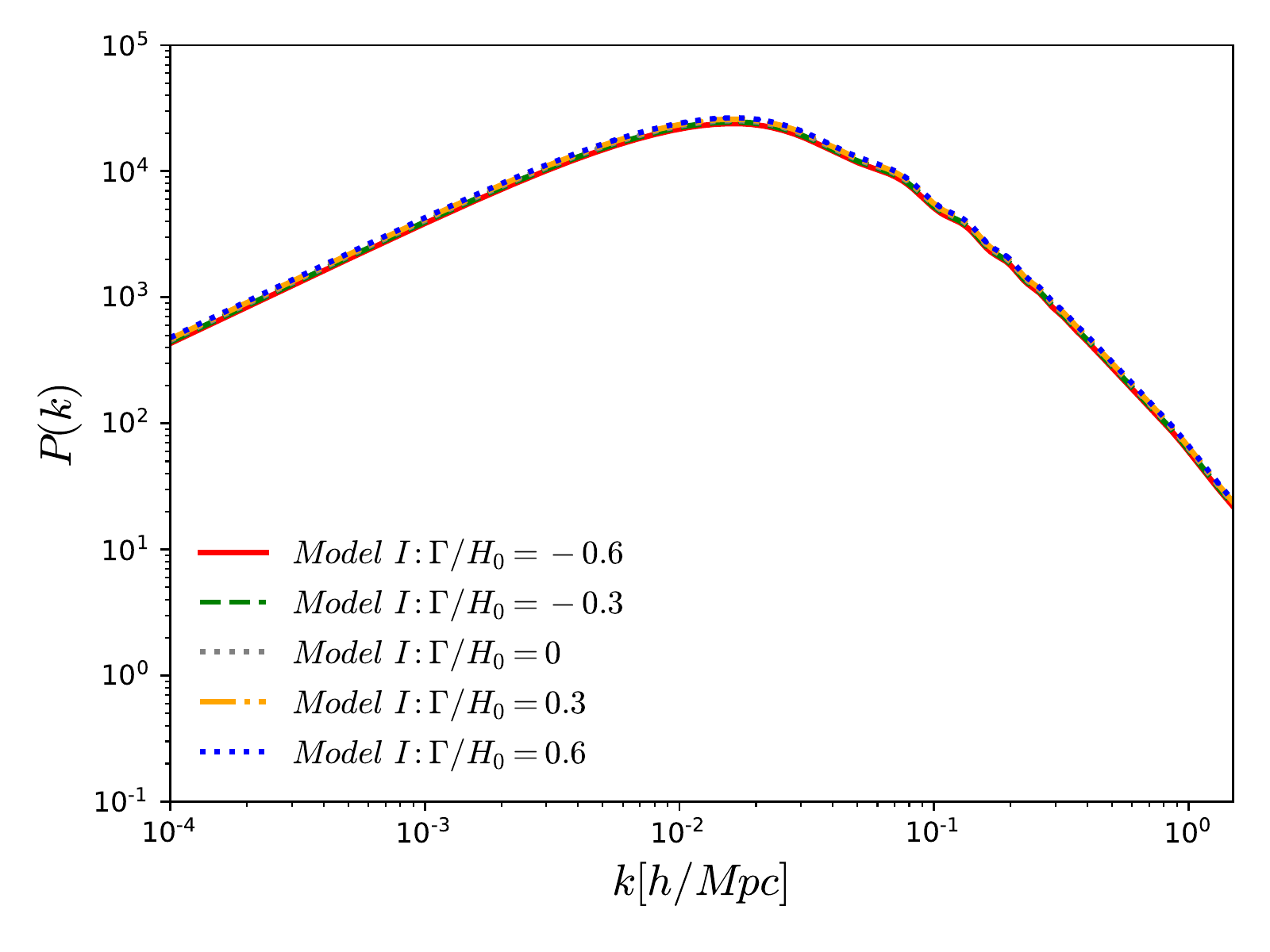}
\caption{CMB temperature angular power spectra (upper left) and matter power spectra (upper right) for different values of the dimensionless parameter $\Gamma/H_0$ of Model I have been shown. }
\label{fig:CMB+matter_modelI}
\end{figure*}

In this context, let us now provide the temperature anisotropies of the CMB spectra and the matter power spectra of Model I.  In Fig.~\ref{fig:CMB+matter_modelI}, we have shown the corresponding plots for various numerical values of the dimensionless parameter $\Gamma/H_0$. In particular, we show the CMB TT spectra in the left panel and matter power spectra in the right one.
One can clearly see that even if we increase the magnitude of $\Gamma/H_0$, there is no significant changes in the spectra. However, a mild deviation from $\Lambda$CDM ($\Gamma/H_{0}=0$) appears only for low multipoles of the CMB spectra.
\begin{figure*}
\includegraphics[width=0.45\textwidth]{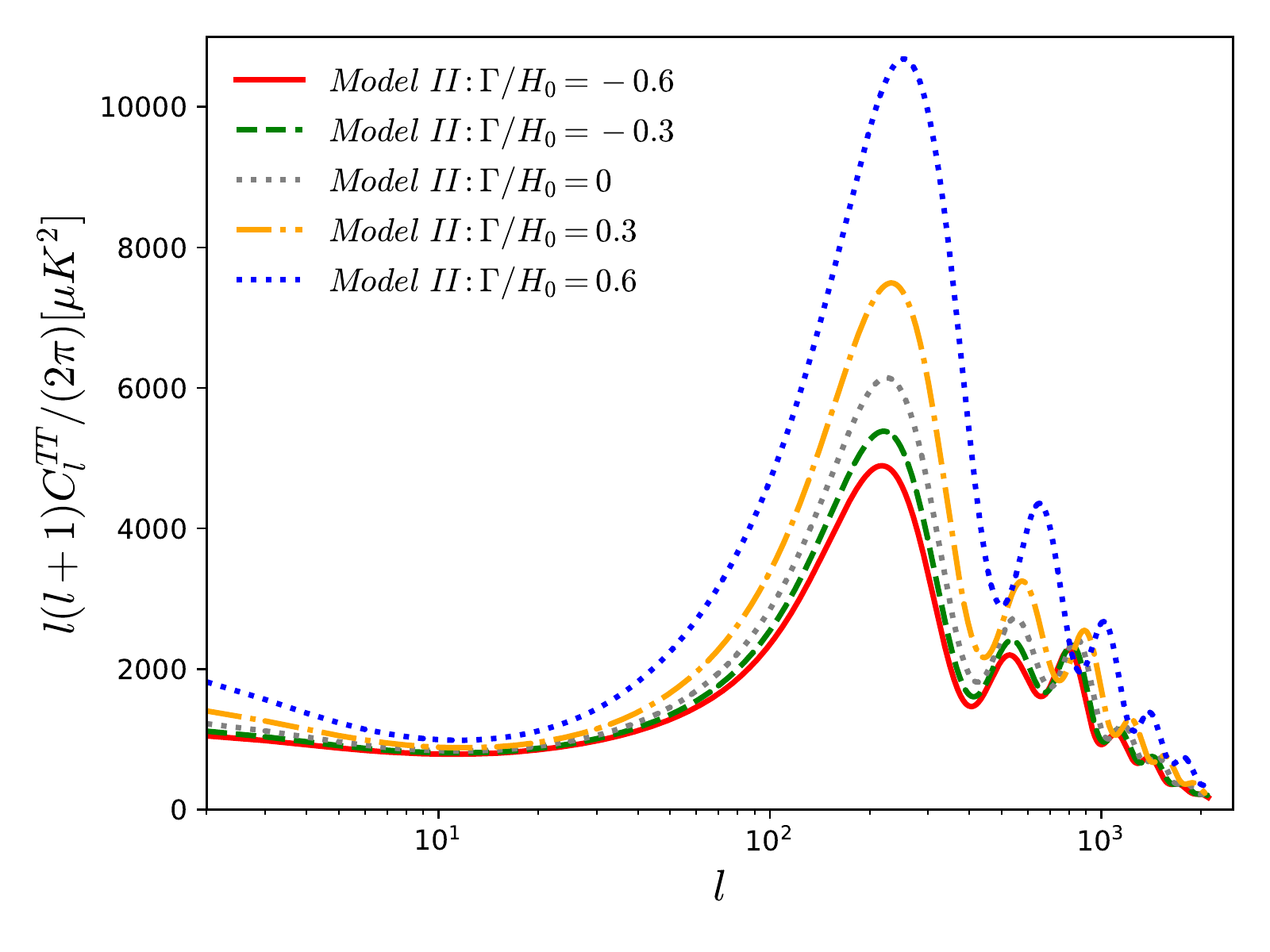}
\includegraphics[width=0.45\textwidth]{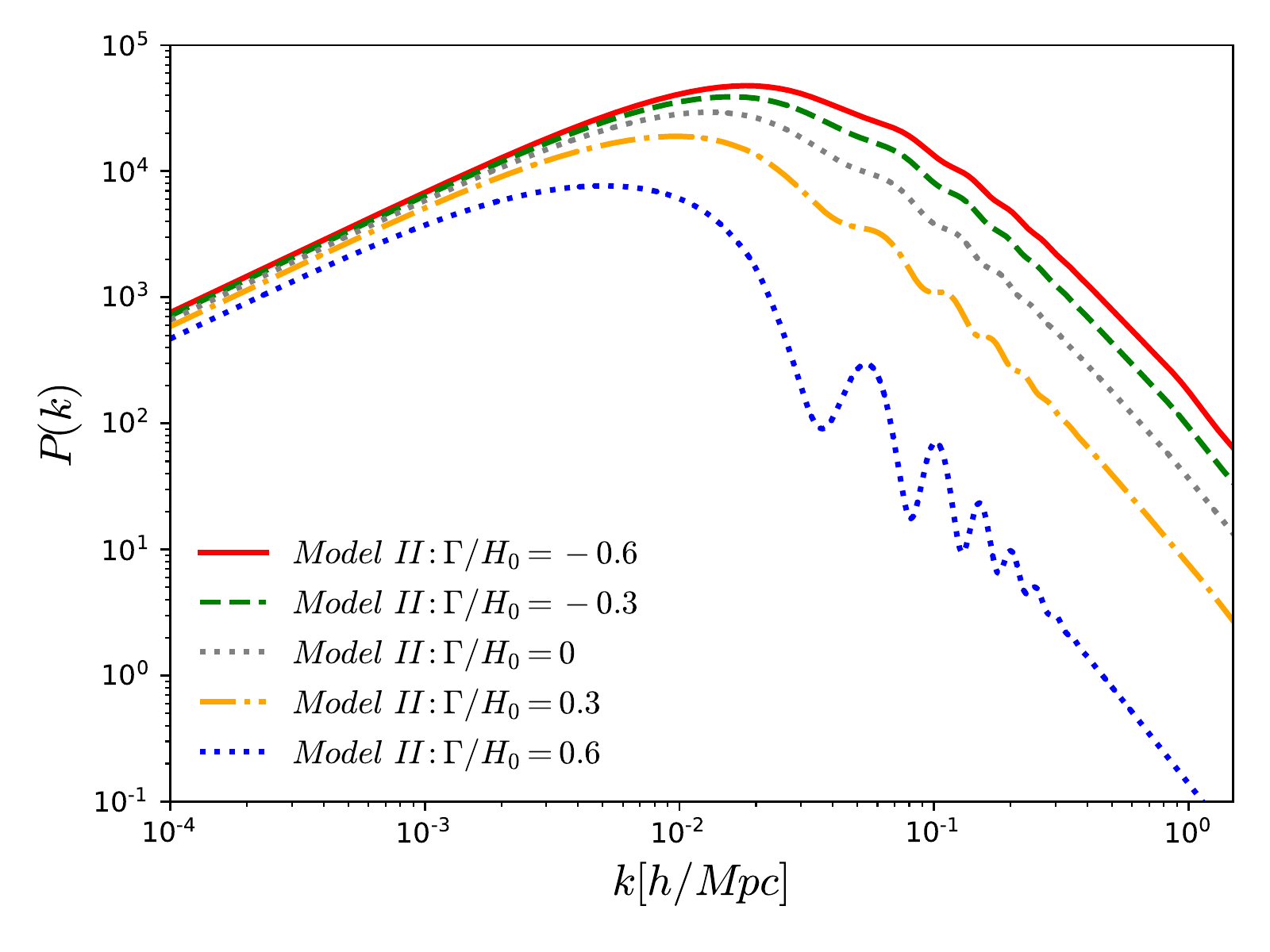}
\caption{CMB temperature angular power spectra (upper left) and matter power spectra (upper right) for different values of the dimensionless coupling parameter $\Gamma/H_0$ of Model II have been shown.}
\label{fig:CMB+matter_modelII}
\end{figure*}

\subsection{Model II}

We now introduce the second metastable DE model in this work which is an  interacting dark scenario between a pressureless dark matter and vacuum energy characterized by the conservation equations:
\begin{eqnarray}
\dot{\rho}_x =  - Q, \label{cons-ide1}\\
\dot{\rho}_c + 3 H \rho_c = Q,\label{cons-ide2}
\end{eqnarray}
where $Q$ refers to an interaction function between these dark sectors. Now, given a specific functional form for $Q$, one may determine the dynamics of the interacting universe by solving the above conservation equations together with the Hubble equation in eqn. (\ref{EFE1}). The possibility of an interaction in the cosmic sector was initially motivated to explain the cosmological constant problem \cite{Wetterich:1994bg} and later this theory was found to provide with an appealing explanation to the cosmic coincidence problem \cite{Amendola:1999er,Cai:2004dk,delCampo:2008sr,delCampo:2008jx}. These results motivated several investigators to work in this region.
Therefore, in the last two decades, cosmological scenarios that allow interaction between the cosmic fluids, namely between the dark sectors of the universe have been extensively studied, see for instance \cite{Zimdahl:2005bk,Wang:2005jx,Berger:2006db,Barrow:2006hia,Sadjadi:2006qb,Bertolami:2007zm,He:2008tn,Valiviita:2008iv,Chen:2008ft,Basilakos:2008ae,Gavela:2009cy,Valiviita:2009nu,Chimento:2009hj,Gavela:2010tm,Harko:2012za,Pan:2013rha,Li:2013bya,Yang:2014vza,Yang:2014gza,Nunes:2014qoa,Faraoni:2014vra,Salvatelli:2014zta,Yang:2014hea,Pan:2012ki,Yin:2015pqa,Li:2015vla,Nunes:2016dlj,Yang:2016evp,Pan:2016ngu,Mukherjee:2016shl,Sharov:2017iue,Shahalam:2017fqt,Guo:2017hea,Cai:2017yww,Yang:2017yme,Yang:2017zjs,Yang:2017ccc,Pan:2017ent,Yang:2018xlt,Yang:2018ubt,Yang:2018pej,vonMarttens:2018iav,Yang:2018qec,Paliathanasis:2019hbi,Barrow:2019jlm,Yang:2019uog}. For these models it has been proposed that  interaction function takes the following forms $Q \propto \rho_c$, $Q \propto \rho_x$, $Q \propto (\rho_c +\rho_x)$, while there are also some other choices which include more complex forms
as far as $Q$  is concerned (see \cite{Yang:2017zjs}).

We would like to stress our original approach regarding the present metastable  model has been phenomenological. Phenomenology is a valid and frequently used method in theoretical cosmology, especially over the last decade.  Indeed a plethora of papers have been published  in metastable dark energy studies, without necessarily providing a  physical interpretation.
Nevertheless,  since Model II allows interactions in the dark sector we would like to point out that there are several attempts regarding the  physical interpretation of
these interactions based on action principles
\cite{vandeBruck:2015ida,Boehmer:2015kta,Boehmer:2015sha,Gleyzes:2015pma,Amico:2016qft,Pan:2020zza}. We remind the reader that in this case cold DM interacts with DE (or vacuum), hence the cold DM density
does not follow the standard power-law $a^{-3}$.

Specifically, it has been found in Ref. \cite{Pan:2020zza} that
the interaction function $Q\propto \rho_{x}$ has a field theoretic description. Moreover, following the recent works  \cite{Basilakos:2019acj,Basilakos:2020qmu}
if we treat $\rho_{x}$ as a running vacuum density $\rho_{\Lambda}(t)$ then
Model II can be seen within the context of a string-inspired
effective theory in the presence of a Kalb-Ramond (KR) gravitational  axion field which descends from the antisymmetric tensor of the  massless gravitational string multiplet.

In the present article, we shall use $Q = \Gamma \rho_x$ as considered in  \cite{Shafieloo:2016bpk,Li:2019san}
where  $\Gamma$ is the coupling parameter. Here we assume that $\Gamma$ is constant
and it has the same dimension as that of the Hubble constant, hence $\Gamma/H_0$ is the dimensionless quantity which we attempt to place constraints from the observational data.
Notice that the present interaction rate does not depend on any parameter related to the expansion of the universe, for instance the Hubble rate of the FLRW universe as considered in many works just for mathematical convenience, and this is the basic feature of the metastable DE models.
The sign of $\Gamma$ determines the flow of energy between the dark two sectors. For $\Gamma >0$, DE decays into DM while for $\Gamma < 0$, the situation is reversed, that means energy flows from DM to DE. We consider a general picture allowing $\Gamma$ to take both positive and negative values, with $\Gamma = 0$ recovering the non-interacting $\Lambda$CDM cosmology. Having presented the gravitational equations for this model at the level of background, one can now proceed towards its understanding at the level of perturbations.

In order to understand the evolution of the model at the level of perturbations,  we recall the perturbed FLRW metric in the synchronous gauge given in
eqn. (\ref{perturbed-metric}).
Within this formalism, one can write down the perturbations equations of the above model as \cite{Wang:2014xca,Yang:2019bpr}:

\begin{eqnarray}
\delta _{c}^{\prime } &=&-\left( \theta _{c}+\frac{h^{\prime }}{2}\right) - \frac{aQ}{\rho_c}\delta_c=-\frac{h^{\prime }}{2}-\left(\frac{a\Gamma\rho_x}{\rho_c}\right)\delta_c, \\
\theta _{c}^{\prime} &=&-\mathcal{H}\theta _{c},  \label{eq:perturbation}
\end{eqnarray}%
where prime denotes the differentiation with respect to the conformal time;
$h$ is the trace of the metric perturbations $h_{ij}$ (see the perturbed metric (\ref{perturbed-metric})); and $\delta_c$ is the density perturbations for the CDM fluid and $
\theta_c$ is the volume expansion scalar for the CDM fluid. Notice here that, following \cite{Wang:2014xca}, we  consider an energy flow parallel to the four velocity of the CDM fluid. As a result, CDM particles follow geodesics as in $\Lambda$CDM and consequently, the vacuum energy perturbations will vanish in the CDM-comoving frame. Now, from
the residual gauge freedom in the synchronous gauge, one may take $\theta_c = 0$ as we have taken, and hence $\theta _{c}^{\prime}  =0$.

We now proceed towards the understanding of the effects of this model through various quantities. In Fig. \ref{fig:CMB+matter_modelII} we plot the temperature anisotropy of the CMB spectra and the matter power spectra for various numerical values of the dimensionless parameter $\Gamma/H_0$. Specifically, the left panel of Fig. \ref{fig:CMB+matter_modelII} shows the CMB TT power spectra and the right panel of Fig. \ref{fig:CMB+matter_modelII} shows the matter power spectra. The features of the spectra are quite different compared to the Model I. As one can see from the CMB TT power spectra, a mild change in the dimensionless coupling parameter $\Gamma/H_0$  produces an observable change in the spectrum and this clearly distinguishes Model II from Model I (see Fig. \ref{fig:CMB+matter_modelI}). In fact, for negative values of $\Gamma/H_0$ (DM decaying into DE), the amplitude of the first acoustic peak in the CMB TT spectra decreases. The opposite scenario holds when the energy flow takes place from DE to DM ($\Gamma > 0$). Similar effects are observed in the matter power spectra, but in this case when $\Gamma/H_0$ increases, the amplitude of the matter power spectrum becomes more
suppressed.

\section{Observational data and methodology}
\label{sec-data}

This section is devoted to describe the observational datasets, statistical techniques and the priors imposed on various free parameters related to the aforementioned metastable dark energy models, namely, Model I and Model II.

Our baseline dataset is Planck 2018, i.e. the latest cosmic microwave background (CMB) temperature and polarization angular power spectra {\it plikTTTEEE+lowl+lowE} from the final 2018 Planck legacy release~\cite{Aghanim:2018eyx,Aghanim:2018oex,Aghanim:2019ame}. Moreover, we test the robustness of our result by including a few cosmological probes, choosing a subset between all the datasets available in the literature (see for example~\cite{Park:2017xbl}):

\begin{itemize}

\item {\bf BAO:} Measurements of the BAO data from different astronomical missions \cite{Beutler:2011hx,Ross:2014qpa,Alam:2016hwk} have been used.

\item {\bf DES:} The galaxy clustering and cosmic shear measurements from the Dark Energy Survey (DES) combined-probe Year 1 results~\cite{Troxel:2017xyo, Abbott:2017wau, Krause:2017ekm}, as adopted by the Planck collaboration in~\cite{Aghanim:2018eyx} have been analyzed.

\item {\bf R19:} The recent measurement of the Hubble constant from a reanalysis of the Hubble Space Telescope data using Cepheids as calibrators, giving $H_0 = 74.03 \pm 1.42$ km/s/Mpc at $68\%$ CL~\cite{Riess:2019cxk} has been considered. It is important to comment that this $H_0$ value is in tension at $4.4 \sigma$ with the Planck's estimation within the $\Lambda$CDM cosmological set-up.

\end{itemize}

To constrain the metastable DE scenarios we use our modified version of the publicly available markov chain monte carlo package \texttt{CosmoMC} \cite{Lewis:2002ah,Lewis:1999bs}, an excellent cosmological code having a fine convergence diagnostic by Gelman-Rubin \cite{Gelman-Rubin}. This code includes the support for Planck 2018 likelihood~\cite{Aghanim:2018oex,Aghanim:2019ame}. The models we are considering have one extra free parameter, $\Gamma$, compared to the flat $\Lambda$CDM model (six-parameters). Let us also mention that in the current analysis, we have fixed the sound speed of DE to unity ($c_{sx}^2 = 1$), which means that
we are dealing with a homogeneous DE.
Therefore, the parameter space of the models is:
\begin{align}
\mathcal{P}_1 \equiv\Bigl\{\Omega_{b}h^2, \Omega_{c}h^2, 100\theta_{MC}, \tau, n_{s}, {\rm log}[10^{10}A_{s}], \Gamma/H_0 \Bigr\}~,
\label{eq:parameter_space1}
\end{align}
where $\Omega_{b} h^2$, $\Omega_{c}h^2$, are the dimensionless densities of baryons and cold dark matter, respectively; $\theta_{MC}$ denotes the ratio of the sound horizon to the angular diameter distance; $\tau$ refers to the reionization optical depth; $n_{s}$ denotes the scalar spectral index; $A_s$ being the amplitude of the primordial scalar power spectrum; and $\Gamma/H_0$ being the free parameter of the metastable models normalized to the Hubble constant value. For the statistical analyses, we have imposed flat priors (see Table \ref{tab:priors}) on the above free parameters.

\begin{table}
\begin{center}
\renewcommand{\arraystretch}{1.2}
\begin{tabular}{|c@{\hspace{0.5 cm}}|@{\hspace{0.25 cm}} c@{\hspace{0.25 cm}}|@{\hspace{0.5 cm}} c|}
\hline
\textbf{Parameter}                    & \textbf{Prior (Model I)}&\textbf{Prior (Model II)}\\
\hline\hline
$\Omega_{b} h^2$             & $[0.005,0.1]$& $[0.005,0.1]$\\
$\Omega_{c} h^2$             & $[0.01,0.99]$& $[0.01,0.99]$\\
$\tau$                       & $[0.01,0.8]$& $[0.01,0.8]$\\
$n_s$                        & $[0.5, 1.5]$& $[0.5, 1.5]$\\
$\log[10^{10}A_{s}]$         & $[2.4,4]$& $[2.4,4]$\\
$100\theta_{MC}$             & $[0.5,10]$& $[0.5,10]$\\
$\Gamma/H_0$                 &  $[-1, 1]$&  $[-1, 0.7]$\\
\hline
\end{tabular}
\end{center}
\caption{We show the flat priors on the free parameters of both metastable DE models for the statistical simulations. }
\label{tab:priors}
\end{table}

\begingroup
\squeezetable
\begin{center}
\begin{table*}
\begin{tabular}{ccccccccccccccc}
\hline\hline
Parameters & Planck 2018 & Planck 2018+BAO & Planck 2018+DES & Planck 2018+R19 & Planck 2018+BAO+DES+R19 \\ \hline

$\Omega_c h^2$ & $    0.1205_{-    0.0014-    0.0027}^{+    0.0014+    0.0027}$ & $    0.1197_{-    0.0012-    0.0024}^{+    0.0013+    0.0024}$ &  $    0.1183_{-    0.0011-    0.0022}^{+    0.0011+    0.0022}$  & $    0.1203_{-    0.0013-    0.0025}^{+    0.0013+    0.0026}$  & $    0.1190_{-    0.00099-    0.0020}^{+    0.00098+    0.0019}$ \\

$\Omega_b h^2$ & $    0.02231_{-    0.00015-    0.00031}^{+    0.00015+    0.00029}$ & $    0.02236_{-    0.00014-    0.00029}^{+    0.00015+    0.00028}$ & $    0.02246_{-    0.00014-    0.00028}^{+    0.00014 +    0.00028}$  & $    0.02232_{-    0.00016-    0.00029}^{+    0.00014+    0.00029}$  & $    0.02243_{-    0.00014-    0.00026}^{+    0.00014+    0.00026}$ \\

$100\theta_{MC}$ & $    1.04062_{-    0.00030-    0.00062}^{+    0.00031+    0.00060}$ & $    1.04072_{-    0.00031-    0.00060}^{+    0.00029+    0.00061}$  & $    1.04084_{-    0.00032-    0.00060}^{+    0.00030+    0.00061}$  & $    1.04065_{-    0.00032-    0.00061}^{+    0.00031+    0.00064}$  & $    1.04077_{-    0.00030-    0.00058}^{+    0.00031+    0.00058}$  \\

$\tau$ & $    0.054_{-    0.0074-    0.015}^{+    0.0074+    0.015}$ & $    0.056_{-    0.0079-    0.016}^{+    0.0077+    0.017}$  & $    0.055_{-    0.0077-    0.016}^{+    0.0077+    0.017}$  & $    0.055_{-    0.0084-    0.015}^{+    0.0077+    0.016}$ & $    0.053_{-    0.0073-    0.015}^{+    0.0073+    0.015}$ \\

$n_s$ & $    0.9722_{-    0.0044-    0.0086}^{+    0.0043+    0.0086}$ & $    0.9740_{-    0.0040-    0.0078}^{+    0.0040+    0.0078}$  & $    0.9766_{-    0.0040-    0.0077}^{+    0.0039+    0.0078}$ & $    0.9729_{-    0.0042-    0.0084}^{+    0.0043+    0.0083}$  & $    0.9750_{-    0.0038-    0.0072}^{+    0.0038+    0.0074}$  \\

${\rm{ln}}(10^{10} A_s)$ & $    3.055_{-    0.015-    0.031}^{+    0.015+    0.031}$ & $    3.056_{-    0.017-    0.033}^{+    0.016+    0.035}$ & $    3.051_{-    0.016-    0.031}^{+    0.016+    0.033}$  & $    3.055_{-    0.017-    0.031}^{+    0.016+    0.032}$ & $    3.048_{-    0.016-    0.029}^{+    0.015+    0.032}$  \\

$\Gamma/H_0$ & $    > 0.04, \, \mbox{unconstrained}$ & $    0.17_{-    0.23-    0.47}^{+    0.26+    0.47}$ & $   >0.54 \, >-0.01$ & $    0.78_{-    0.08}^{+    0.19}\, >0.53$   & $ > 0.367 > 0.193   $  \\

$\Omega_{m0}$ & $    0.303_{-    0.053-    0.065}^{+    0.026+    0.080}$ & $    0.306_{-    0.016-    0.026}^{+    0.014+    0.028}$ & $    0.263_{-    0.027-    0.037}^{+    0.012+    0.048}$ & $    0.263_{-    0.011-    0.019}^{+    0.0089+    0.020}$  & $    0.275_{-    0.0089-    0.017}^{+    0.0076+    0.018}$ \\

$H_0$ & $   69.3_{-    3.5-    8.3}^{+    5.9+    7.3}$ & $   68.3_{-    1.7-    3.4}^{+    1.6+    3.2}$ & $   73.6_{-    1.8-    6.2}^{+    3.7+    4.9}$  & $   73.8_{-    1.2-    2.6}^{+    1.4+    2.5}$  & $   71.94_{-    1.08-    2.42}^{+    1.08+    2.21}$ \\

\hline
$\chi^2$ & 2771.046 & 2779.456 & 3293.906 & 2771.620 & 3313.11 \\
\hline\hline
\end{tabular}
\caption{Summary of the observational constraints  and lower limits at 68\% and 95\% CL on the cosmological scenario driven by the metastable DE scenario,  {\it Model I}, using different observational datasets. The parameters are varying in the ranges described in Table~\ref{tab:priors}. }
\label{tab:M1}
\end{table*}
\end{center}
\endgroup
\begin{figure*}
\includegraphics[width=0.7\textwidth]{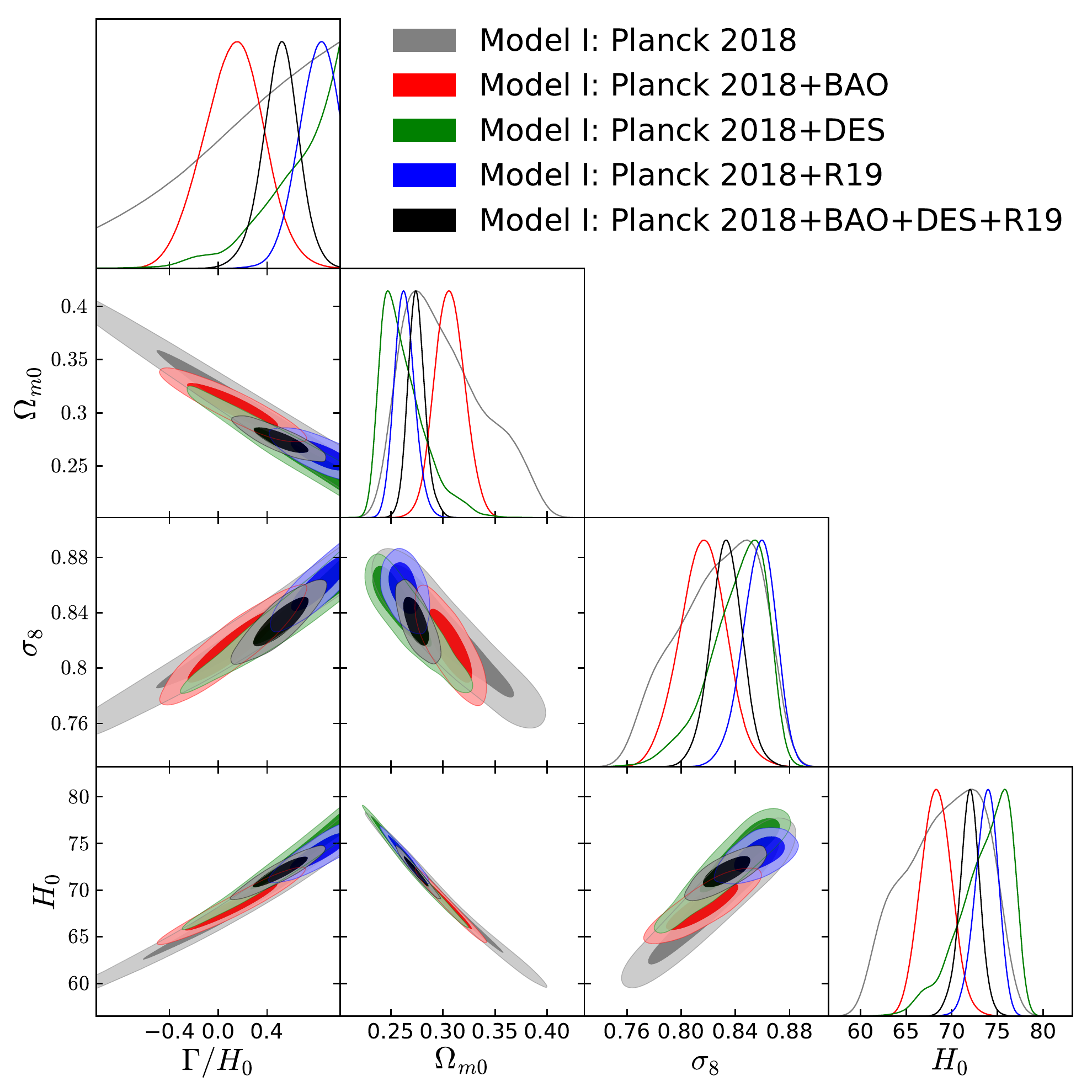}
\caption{68\% and 95\% CL constraints on the metastable DE scenario, {\it Model I}, using various observational datasets have been displayed. }
\label{fig:model1}
\end{figure*}
\begin{figure*}
\includegraphics[width=0.45\textwidth]{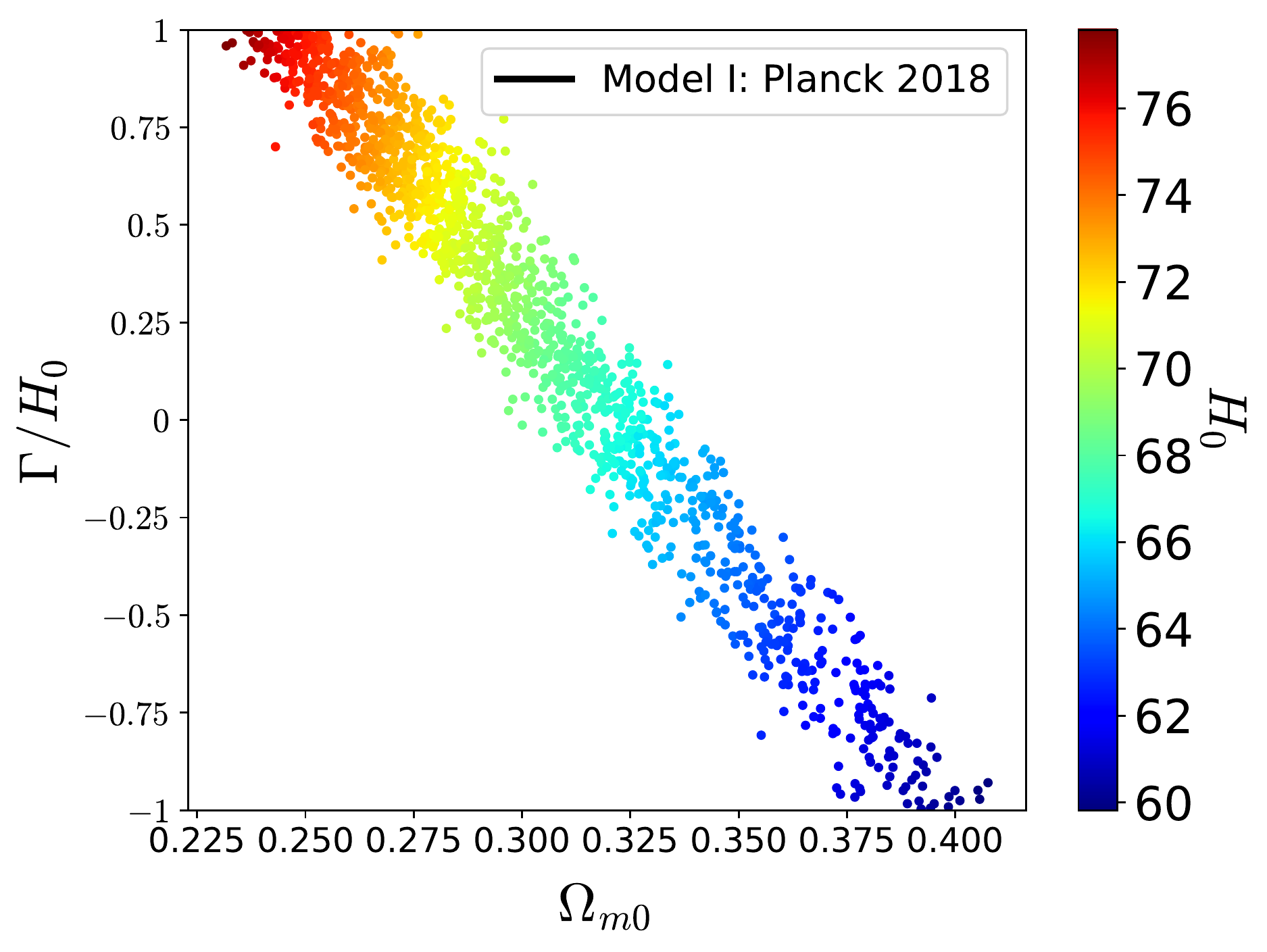}
\includegraphics[width=0.45\textwidth]{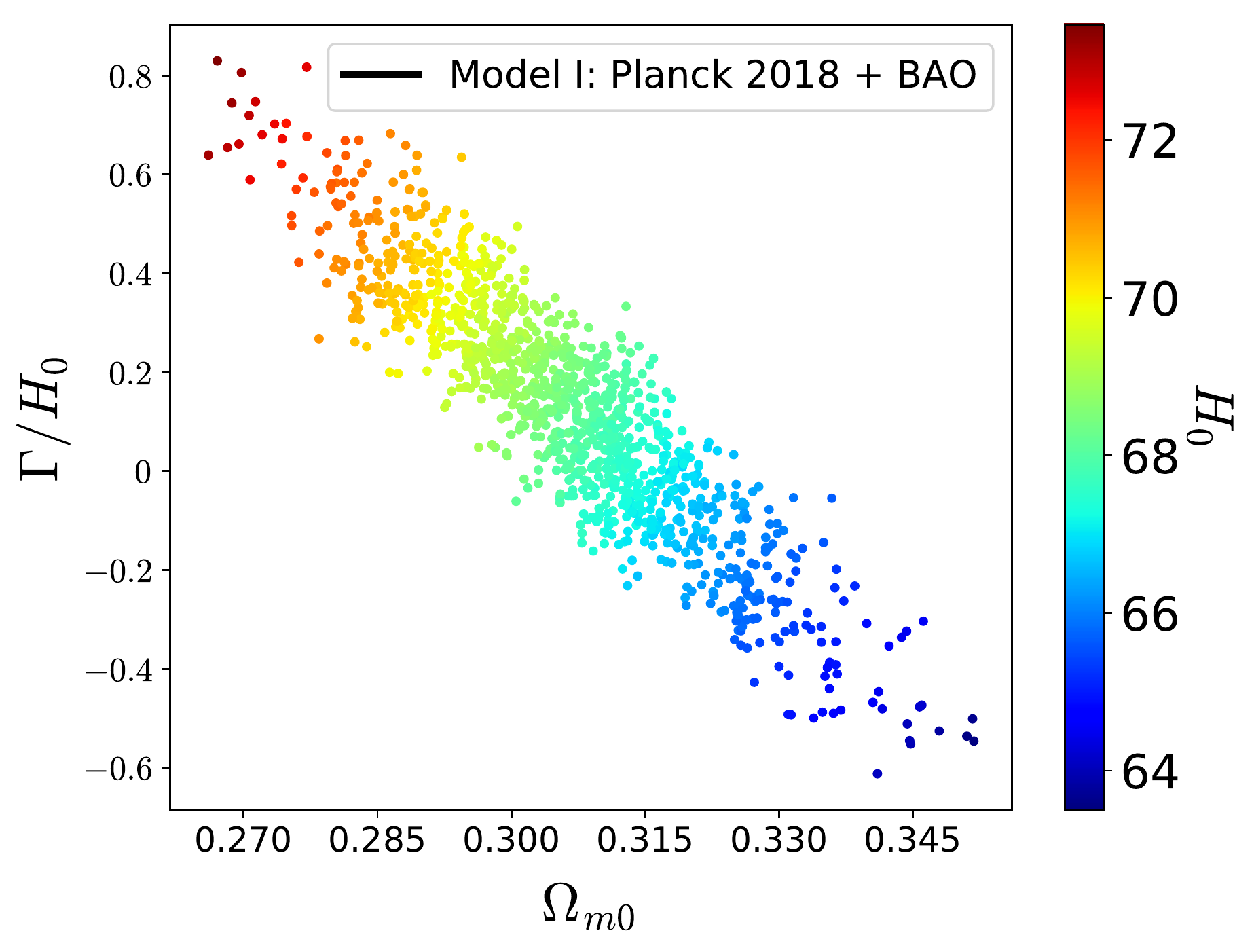}\\
\includegraphics[width=0.45\textwidth]{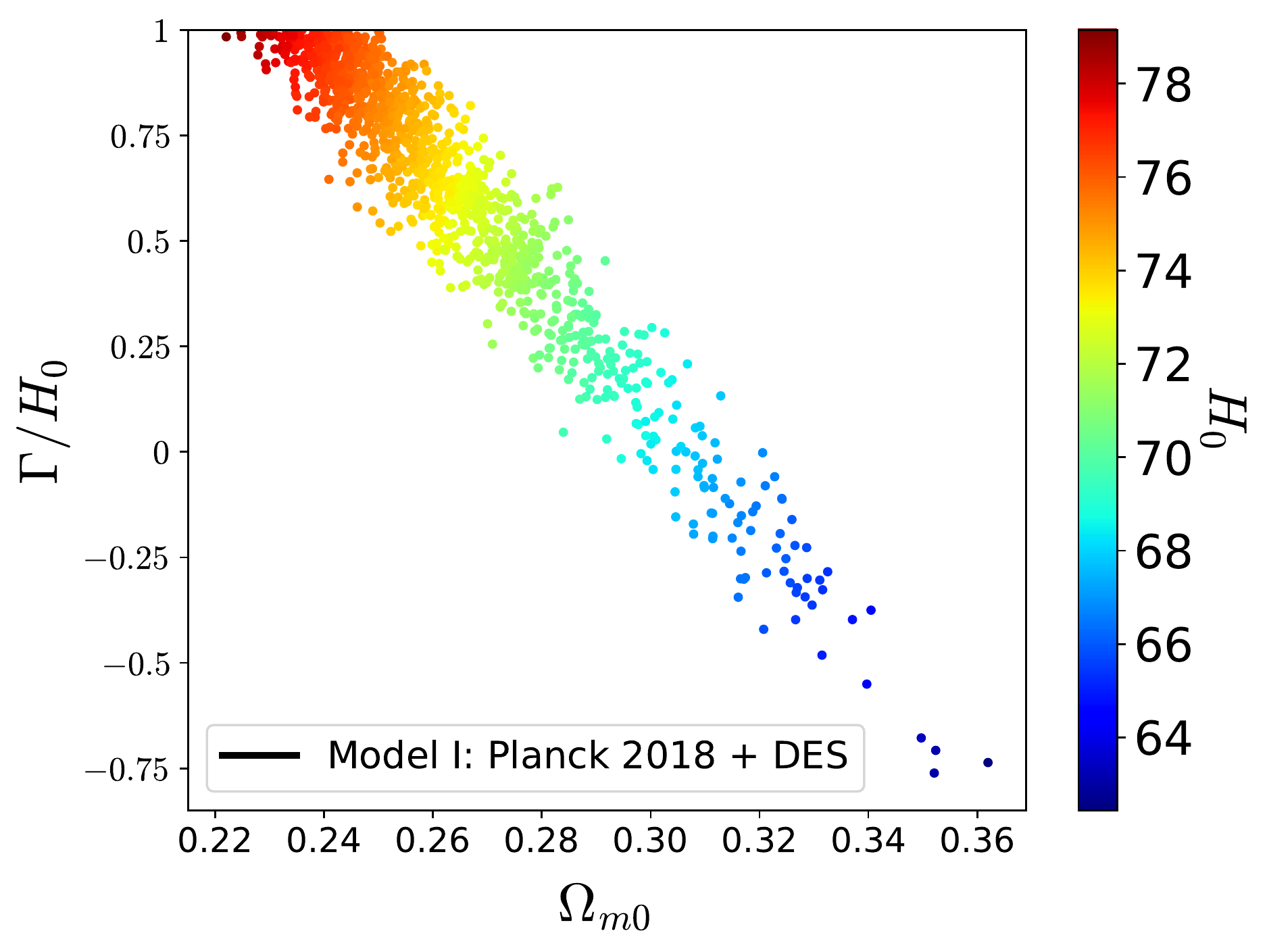}
\includegraphics[width=0.45\textwidth]{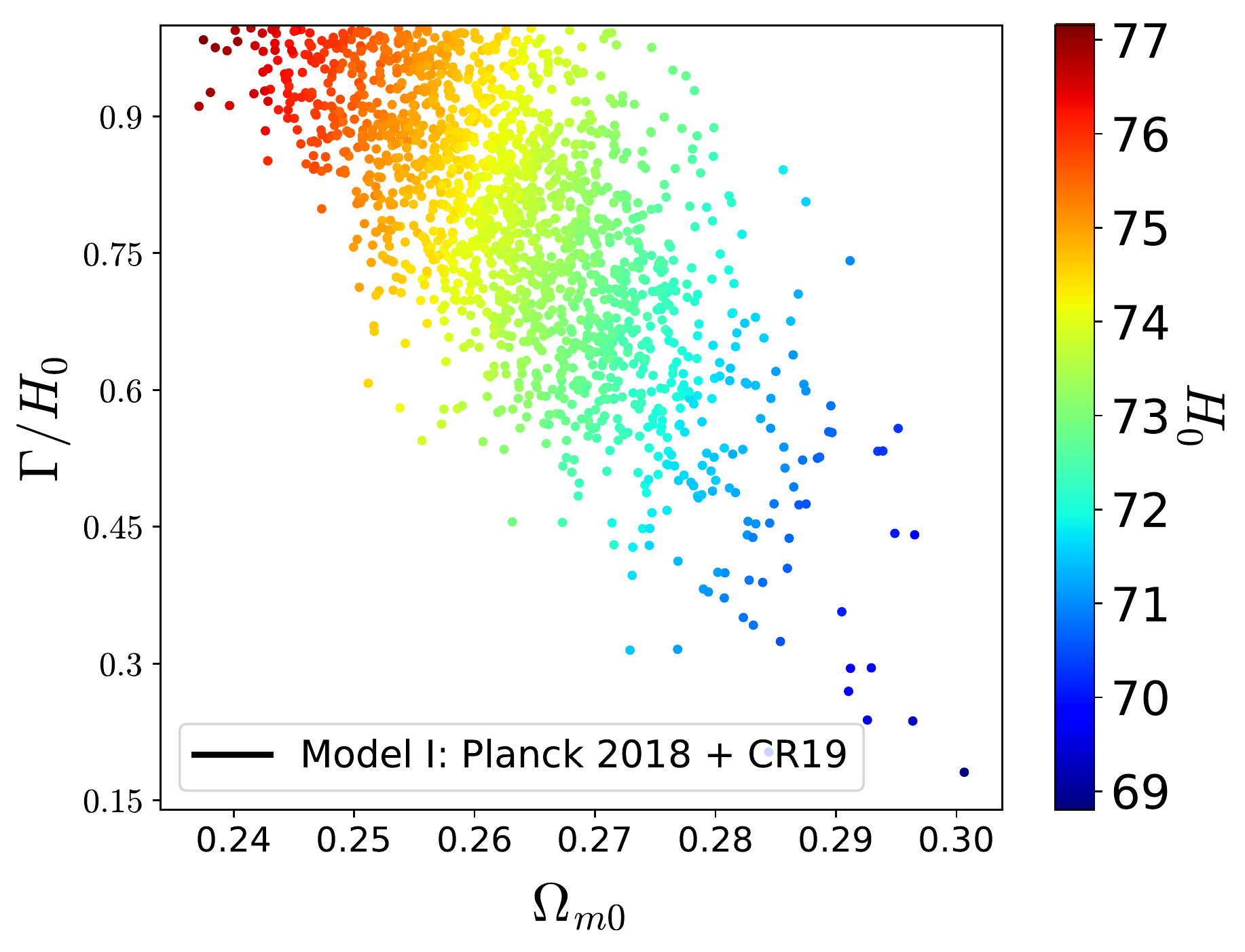}\\
\includegraphics[width=0.45\textwidth]{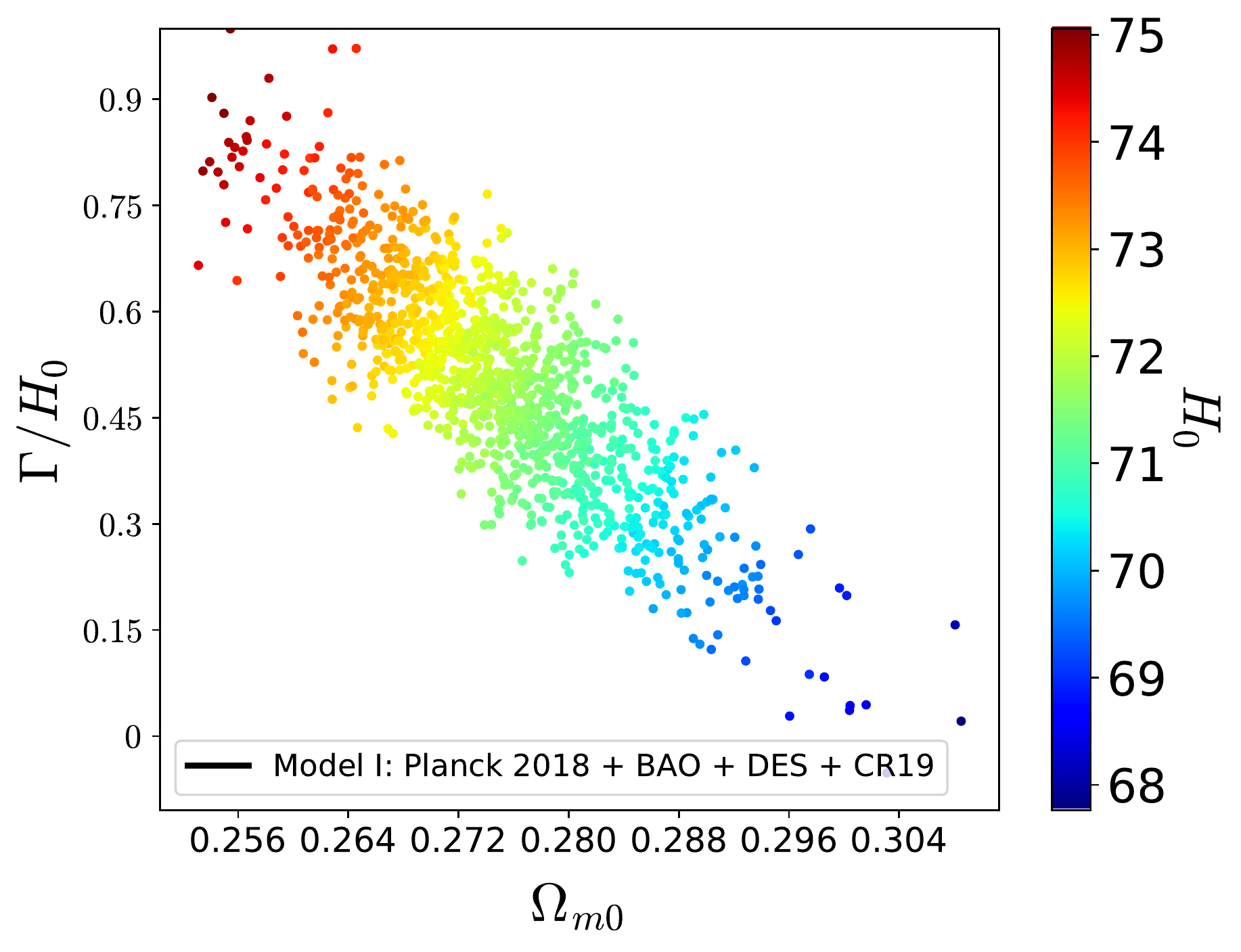}
\caption{3D scattered plots at 95\% CL in the plane $\Gamma/H_0$ vs $\Omega_{m0}$, coloured by the Hubble constant value $H_0$ for Model I. A strong anti-correlation between $\Gamma/H_0$ and $\Omega_{m0}$, and a positive correlation between $\Gamma/H_0$ and $H_0$ are present. For Planck alone, upper left panel, $\Gamma/H_0$ is unconstrained, while the addition of external datasets to Planck 2018 helps in constraining this parameter. }
\label{fig-3d-model1}
\end{figure*}

\begin{figure*}
\includegraphics[width=0.7\textwidth]{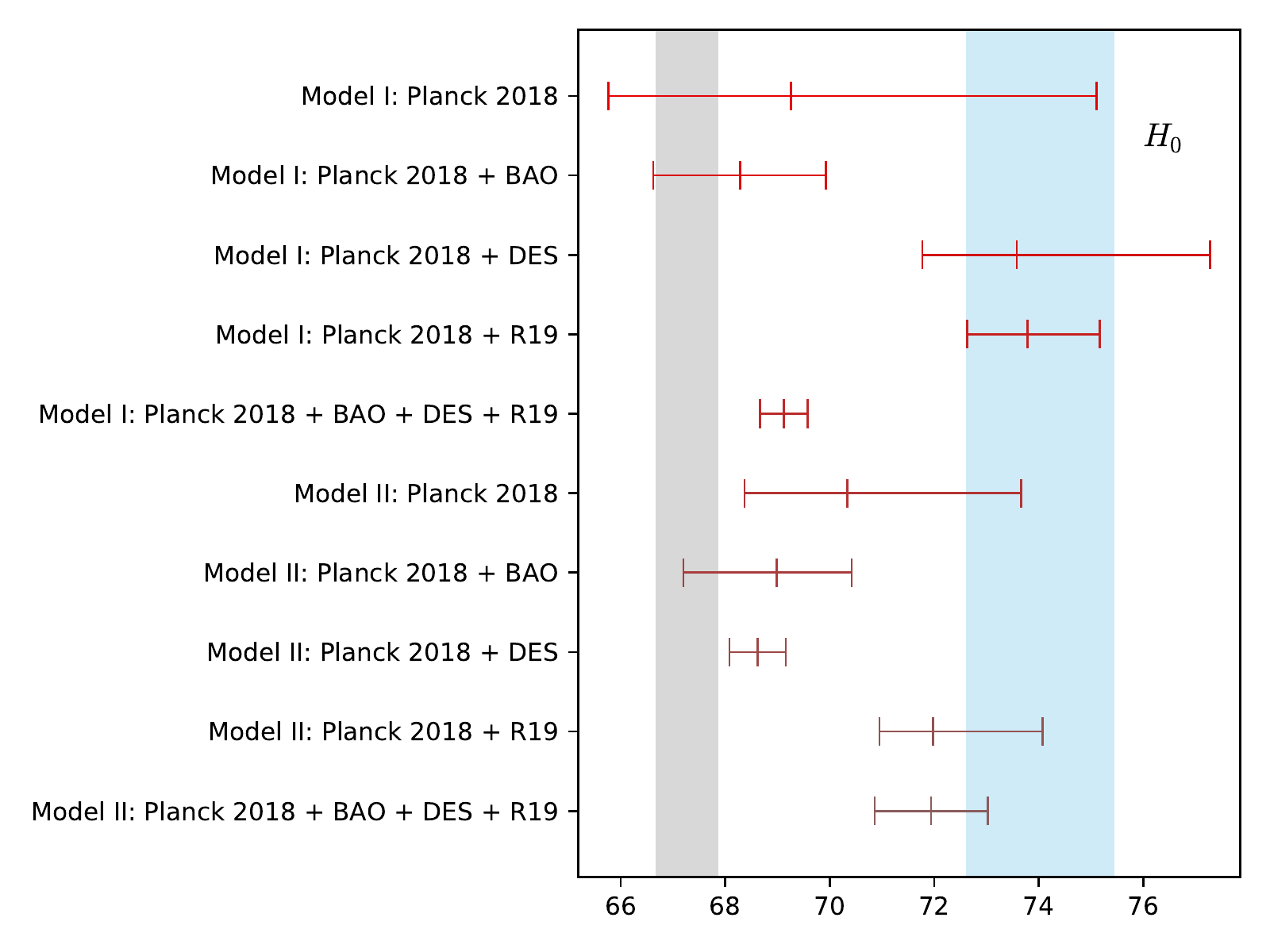}
\caption{Whisker plot with 68\%~CL constraints on $H_0$ for the metastable DE models (Model I and Model II) for various observational datasets use here. The grey vertical band corresponds to the estimation of $H_0$ by the final Planck 2018 release~\cite{Aghanim:2018eyx} and the sky blue vertical band corresponds to the R19 value of $H_0$, as measured by the SH0ES collaboration in~\cite{Riess:2019cxk}. }
\label{fig:whisker}
\end{figure*}

\begingroup
\squeezetable
\begin{center}
\begin{table*}
\begin{tabular}{ccccccccccccc}
\hline\hline
Parameters & Planck 2018 & Planck 2018+BAO & Planck 2018+DES & Planck 2018+R19 & Planck 2018+BAO+DES+R19  \\ \hline

$\Omega_c h^2$ & $    0.1202_{-    0.0014-    0.0026}^{+    0.0014+    0.0027}$ & $    0.1193_{-    0.0010-    0.0020}^{+    0.0010+    0.0019}$ & $    0.1179_{-    0.0010-    0.0021}^{+    0.0010+    0.0021}$     &  $    0.1179_{-    0.0012-    0.0025}^{+    0.0012+    0.0025}$   & $    0.1172_{-    0.00094-    0.0016}^{+    0.00084+    0.0017}$  \\

$\Omega_b h^2$ & $    0.02236_{-    0.00015-    0.00028}^{+    0.00015+    0.00029}$ & $    0.02243_{-    0.00014-    0.00027}^{+    0.00014+    0.00027}$ &    $    0.02251_{-    0.00014-    0.00026}^{+    0.00014+    0.00027}$  & $    0.02255_{-    0.00014-    0.00028}^{+    0.00014+    0.00028}$   & $    0.02260_{-    0.00012-    0.00026}^{+    0.00013+    0.00025}$ \\

$100\theta_{MC}$ & $    1.04091_{-    0.00031-    0.00061}^{+    0.00030+    0.00061}$ & $    1.04100_{-    0.00029-    0.00058}^{+    0.00029+    0.00057}$ & $    1.04113_{-    0.00030-    0.00059}^{+    0.00030+    0.00060}$   &$    1.04120_{-    0.00030-    0.00059}^{+    0.00030+    0.00057}$   & $    1.04125_{-    0.00029-    0.00056}^{+    0.00029+    0.00055}$  \\

$\tau$ & $    0.054_{-    0.0083-    0.015}^{+    0.0071+    0.016}$ & $    0.055_{-    0.0084-    0.015}^{+    0.0076+    0.017}$ & $    0.055_{-    0.0081-    0.015}^{+    0.0072+    0.016}$   & $    0.058_{-    0.0085-    0.016}^{+    0.0075+    0.016}$    & $    0.056_{-    0.0073-    0.015}^{+    0.0071+    0.015}$  \\

$n_s$ & $    0.9647_{-    0.0043-    0.0084}^{+    0.0044+    0.0085}$ & $    0.9669_{-    0.0038-    0.0073}^{+    0.0038+    0.0075}$  & $    0.9694_{-    0.0039-    0.0078}^{+    0.0039+    0.0078}$  & $    0.9704_{-    0.0041-    0.0083}^{+    0.0041+    0.0082}$  & $    0.9715_{-    0.0036-    0.0072}^{+    0.0035+    0.0072}$ \\

${\rm{ln}}(10^{10} A_s)$ & $    3.045_{-    0.017-    0.030}^{+    0.015+    0.032}$ & $    3.045_{-    0.016-    0.032}^{+    0.016+    0.034}$ & $    3.039_{-    0.017-    0.030}^{+    0.015+    0.032}$  & $    3.047_{-    0.017-    0.034}^{+    0.016+    0.033}$   & $    3.042_{-    0.015-    0.028}^{+    0.015+    0.030}$  \\

$\Omega_{m0}$ & $    0.317_{-    0.0084-    0.016}^{+    0.0084+    0.017}$ & $    0.311_{-    0.0060-    0.012}^{+    0.0060+    0.012}$ & $    0.303_{-    0.0061-    0.012}^{+    0.0061+    0.012}$  & $    0.302_{-    0.0073-    0.014}^{+    0.0073+    0.015}$   & $    0.298_{-    0.0054-    0.0092}^{+    0.0048+    0.010}$ \\

$H_0$ & $   67.27_{-    0.60-    1.20}^{+    0.61+    1.20}$ & $   67.68_{-    0.44-    0.87}^{+    0.45+    0.91}$ & $   68.28_{-    0.48-    0.91}^{+    0.47+    0.96}$   & $   68.35_{-    0.56-    1.11}^{+    0.55+    1.12}$   & $   68.66_{-    0.38-    0.76}^{+    0.41+    0.73}$  \\

\hline

$\chi^2$ &  2773.168 & 2779.690 & 3294.578  &  2791.542 & 3318.602 \\

\hline\hline

\end{tabular}
\caption{We show the constraints on the $\Lambda$CDM scenario (corresponding to $\Gamma =0$) using the same observational data. }
\label{tab:LCDM}
\end{table*}
\end{center}
\endgroup

\begingroup
\squeezetable
\begin{center}
\begin{table*}
\begin{tabular}{ccccccccccc}
\hline\hline
Parameters & Planck 2018  & Planck 2018+BAO & Planck 2018+DES & Planck 2018+R19 & Planck 2018+BAO+DES+R19 \\ \hline

$\Omega_c h^2$ & $    0.064_{-    0.062}^{+    0.022}\, <0.134$ & $    0.091_{-    0.023-    0.056}^{+    0.034+    0.051}$ & $    0.0998_{-    0.0077-    0.014}^{+    0.0071+    0.015}$ &  $    <0.050 \, <0.099$  & $    0.0983_{-    0.0090-    0.0142}^{+    0.0079+    0.0153}$  \\

$\Omega_b h^2$ & $    0.02231_{-    0.00015-    0.00031}^{+    0.00015+    0.00030}$ &   $    0.02233_{-    0.00014-    0.00028}^{+    0.00014+    0.00028}$ &  $    0.02237_{-    0.00015-    0.00029}^{+    0.00015+    0.00029}$ &  $    0.02236_{-    0.00016-    0.00028}^{+    0.00014+    0.00030}$  & $    0.02246_{-    0.00013-    0.00026}^{+    0.00013+    0.00026}$ \\

$100\theta_{MC}$ & $    1.0444_{-    0.0033-    0.0049}^{+    0.0031+    0.0049}$ &  $    1.0425_{-    0.0022-    0.0032}^{+    0.0012+    0.0037}$ & $    1.04183_{-    0.00049-    0.00101}^{+    0.00050+    0.00095}$ & $    1.0461_{-    0.0017-    0.0046}^{+    0.0031+    0.0039}$  & $    1.04202_{-    0.00052-    0.00101}^{+    0.00057+    0.00101}$  \\

$\tau$ & $    0.054_{-    0.0077-    0.015}^{+    0.0075+    0.016}$ & $    0.055_{-    0.0081-    0.015}^{+    0.0076+    0.016}$  &  $    0.055_{-    0.0076-    0.016}^{+    0.0077+    0.016}$ &  $    0.055_{-    0.0081-    0.015}^{+    0.0071+    0.016}$  & $    0.058_{-    0.0077-    0.015}^{+    0.0074+    0.016}$ \\

$n_s$ & $    0.9724_{-    0.0042-    0.0081}^{+    0.0040+    0.0082}$ &  $    0.9736_{-    0.0039-    0.0079}^{+    0.0039+    0.0079}$ &  $    0.9739_{-    0.0040-    0.0083}^{+    0.0041+    0.0081}$ &  $    0.9740_{-    0.0041-    0.0082}^{+    0.0041+    0.0083}$  & $    0.9761_{-    0.0037-    0.0071}^{+    0.0038+    0.0068}$  \\

${\rm{ln}}(10^{10} A_s)$ & $    3.055_{-    0.016-    0.033}^{+    0.016+    0.033}$ &  $    3.056_{-    0.016-    0.032}^{+    0.015+    0.032}$ &  $    3.056_{-    0.017-    0.032}^{+    0.015+    0.033}$ & $    3.056_{-    0.015-    0.030}^{+    0.015+    0.032}$  & $    3.059_{-    0.016-    0.031}^{+    0.016+    0.033}$  \\

$\Gamma/H_0$ & $  <-0.39 \, <0.19$ &  $   -0.29_{-    0.28-    0.53}^{+    0.30+    0.54}$ & $ -0.219_{-    0.090-    0.17}^{+    0.082+    0.17}$ & $  <-0.66 \, <-0.21$  & $   -0.219_{-    0.099-    0.160}^{+    0.089+    0.174}$  \\

$\Omega_{m0}$ & $    0.18_{-    0.13-    0.16}^{+    0.07+    0.19}$ & $    0.242_{-    0.063-    0.14}^{+    0.079+    0.13}$  &  $    0.261_{-    0.019-    0.034}^{+    0.017+    0.038}$ &  $    0.127_{-    0.084-    0.098}^{+    0.031+    0.140}$  & $    0.254_{-    0.023-    0.035}^{+    0.018+    0.038}$  \\

$H_0$ & $   70.3_{-    2.0-    4.9}^{+    3.3+    4.3}$ & $   69.0_{-    1.8-    3.0}^{+    1.4+    3.1}$ & $   68.62_{-    0.54-    1.1}^{+    0.54+    1.1}$ & $   72.0_{-    1.0-    3.4}^{+    2.1+    2.7}$  & $   69.12_{-    0.45-    0.86}^{+    0.46+    0.83}$ \\

\hline

$\chi^2$ & 2771.716 & 2780.014 & 3295.094 & 2775.360 & 3315.868 \\

\hline\hline
\end{tabular}
\caption{Summary of the observational constraints and upper limits at 68\% and 95\% CL on the cosmological scenario driven by the metastable DE scenario, {\it Model II}, using different observational datasets. The parameters are varying in the ranges described in Table~\ref{tab:priors}.}
\label{tab:M2}
\end{table*}
\end{center}
\endgroup
\begin{figure*}
\includegraphics[width=0.7\textwidth]{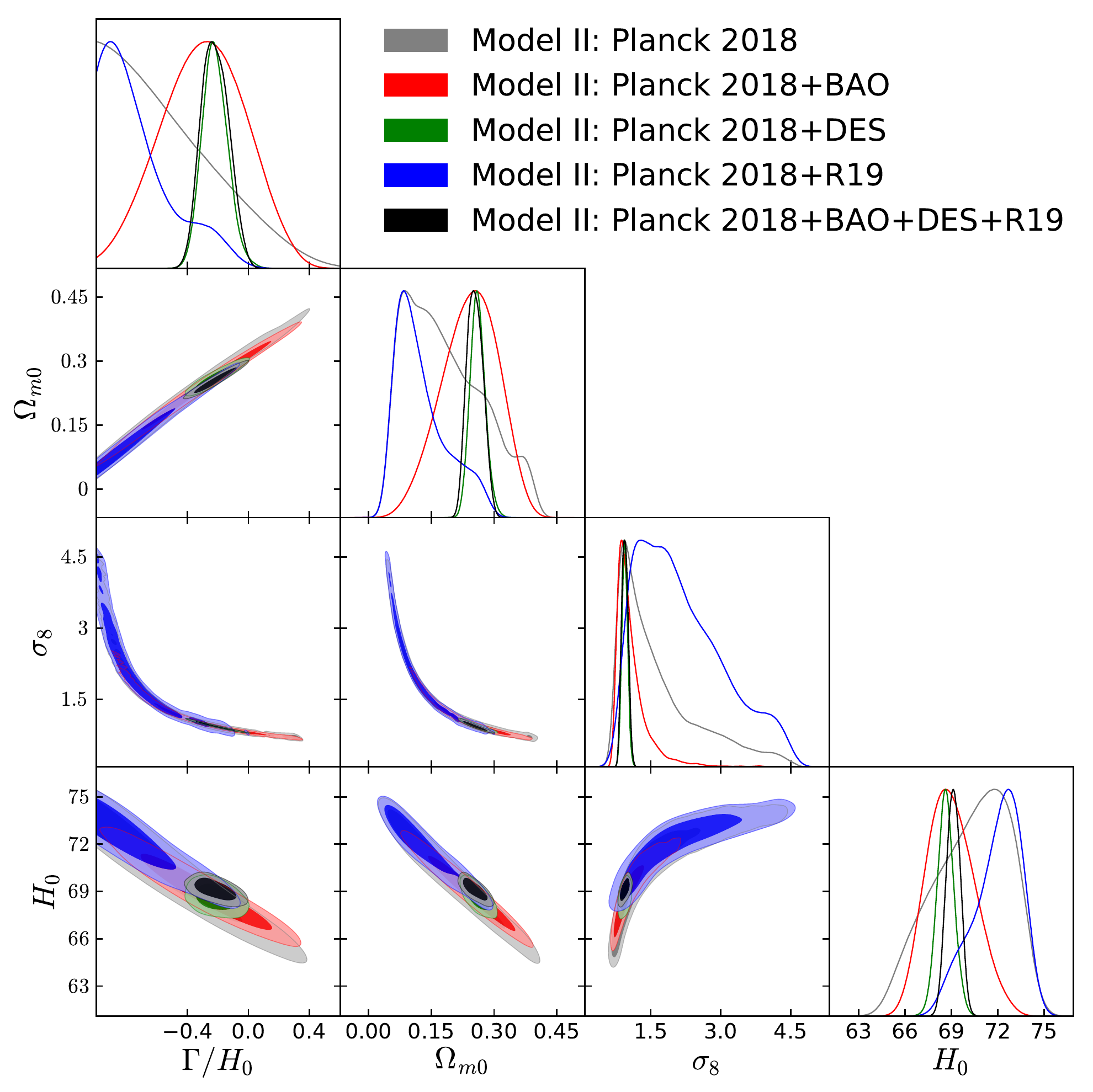}
\caption{68\% and 95\% CL constraints on the metastable DE scenario,  {\it Model II}, using various observational datasets have been displayed. }
\label{fig:model2}
\end{figure*}

\begin{figure*}
\includegraphics[width=0.45\textwidth]{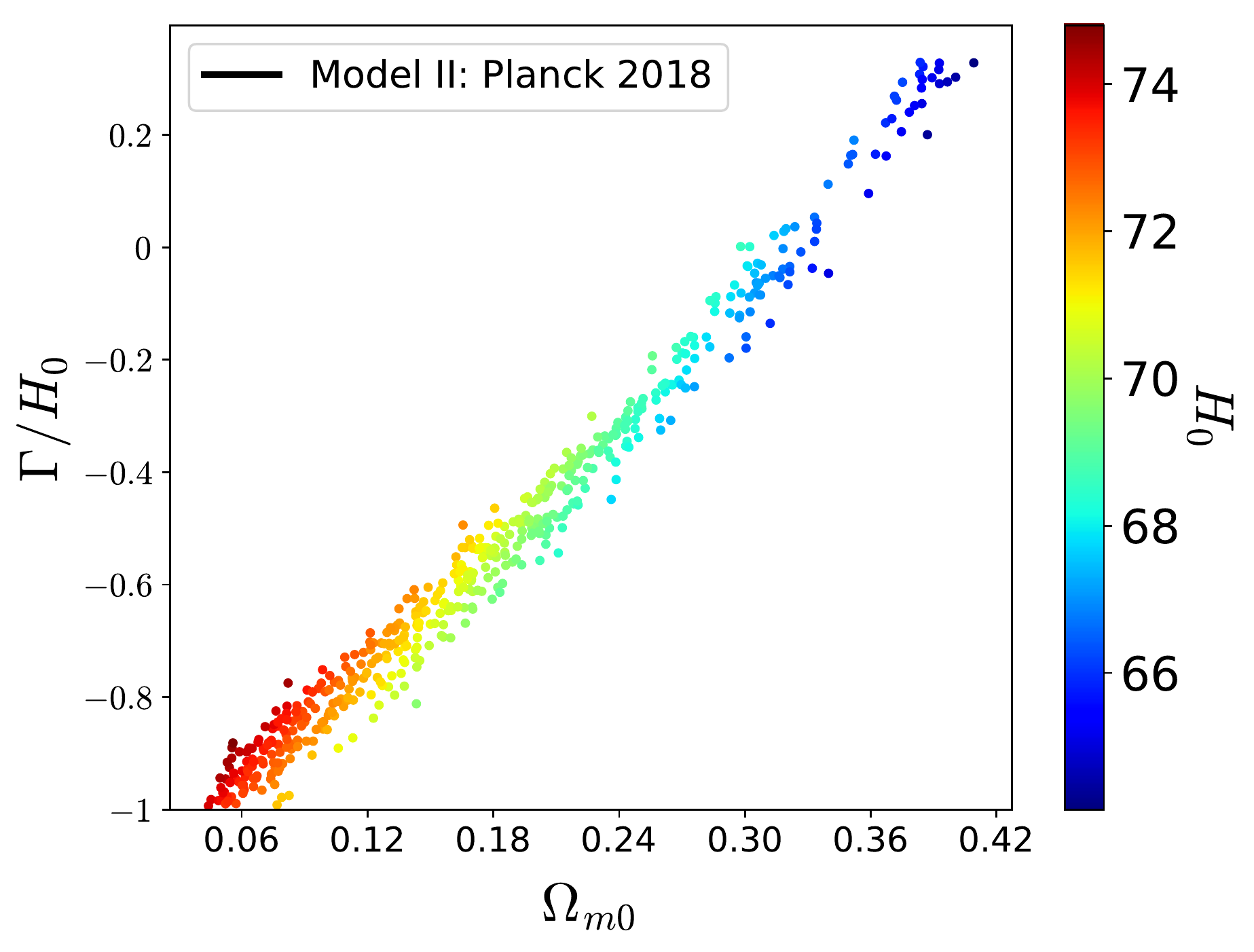}
\includegraphics[width=0.45\textwidth]{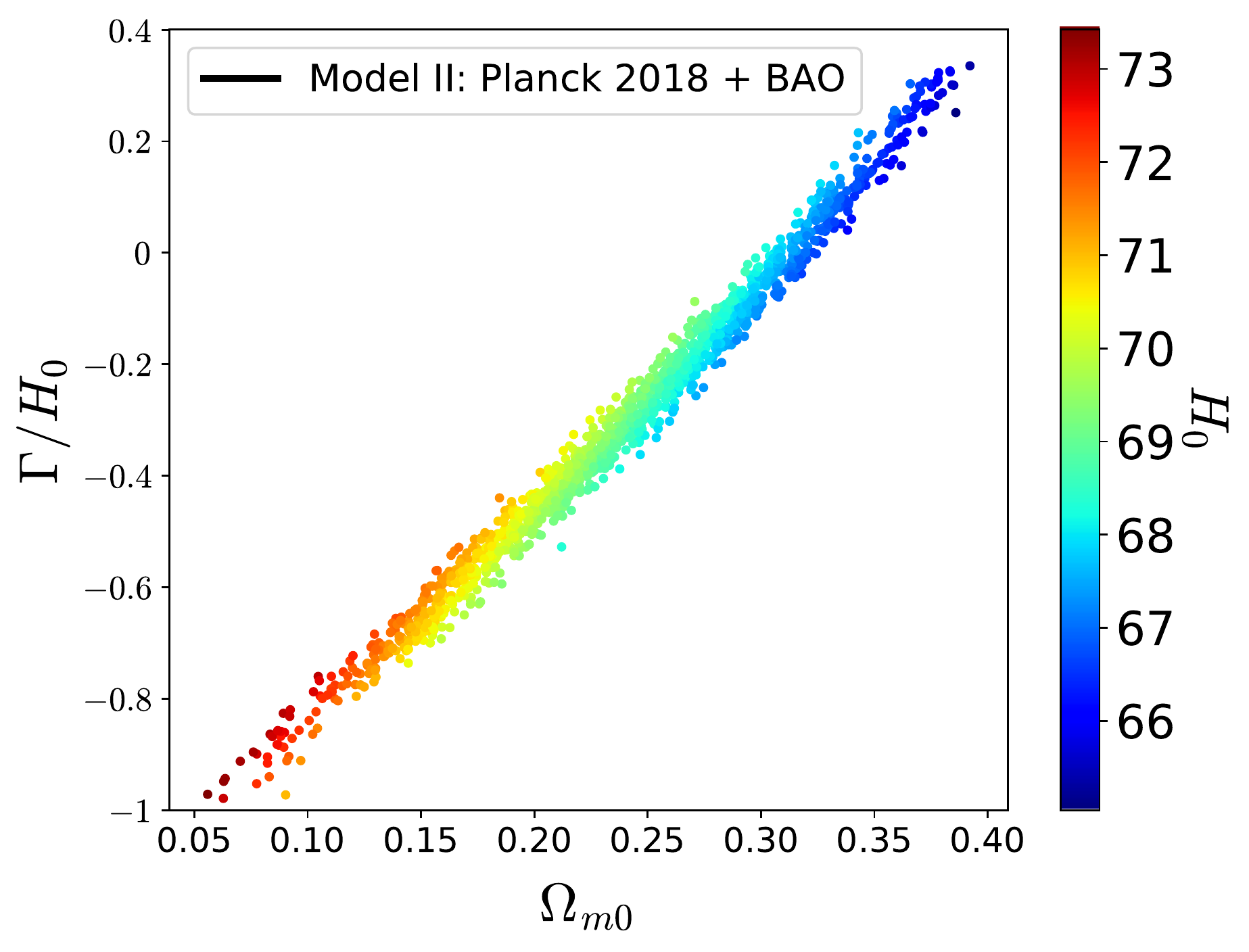}\\
\includegraphics[width=0.45\textwidth]{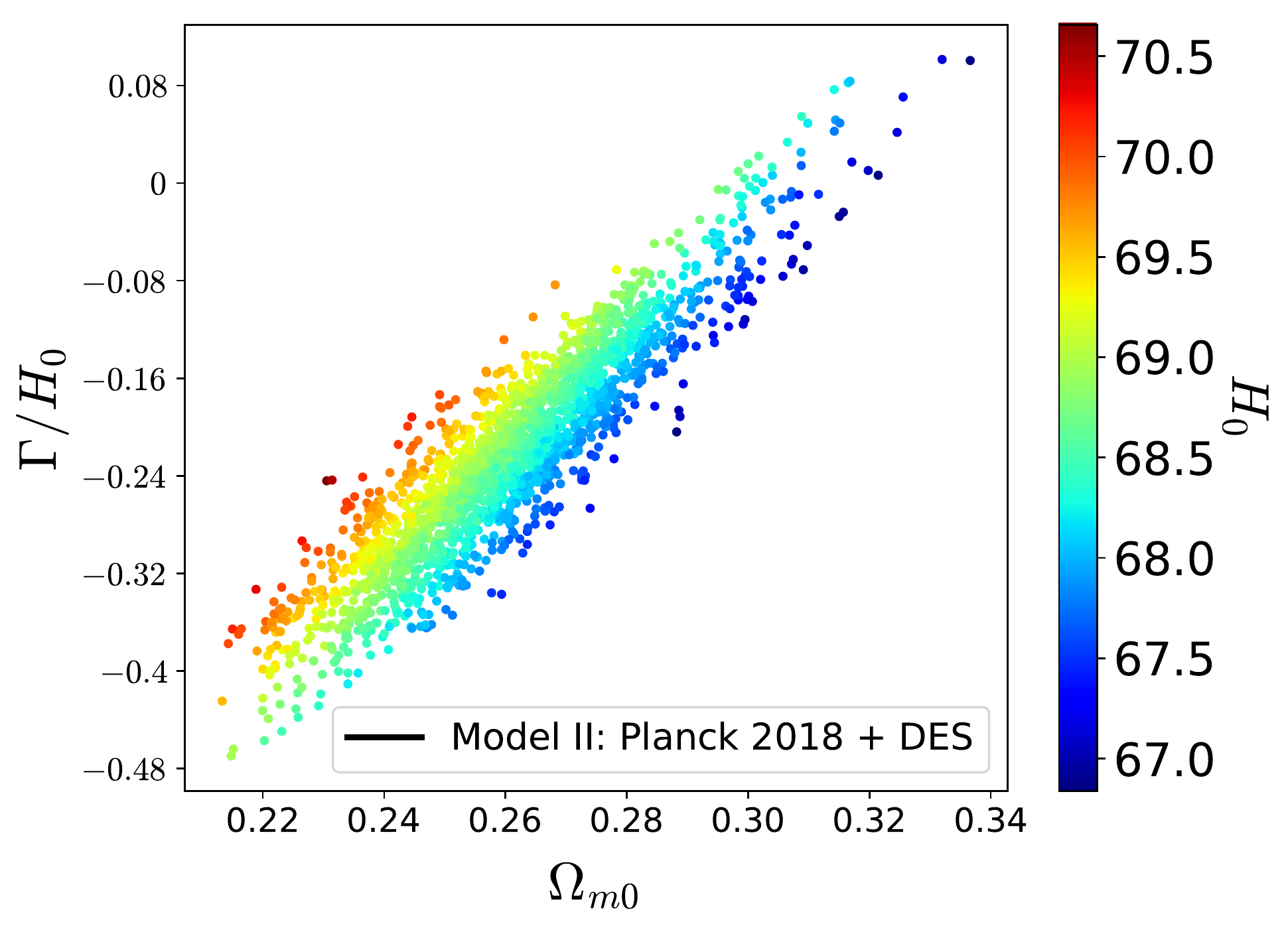}
\includegraphics[width=0.45\textwidth]{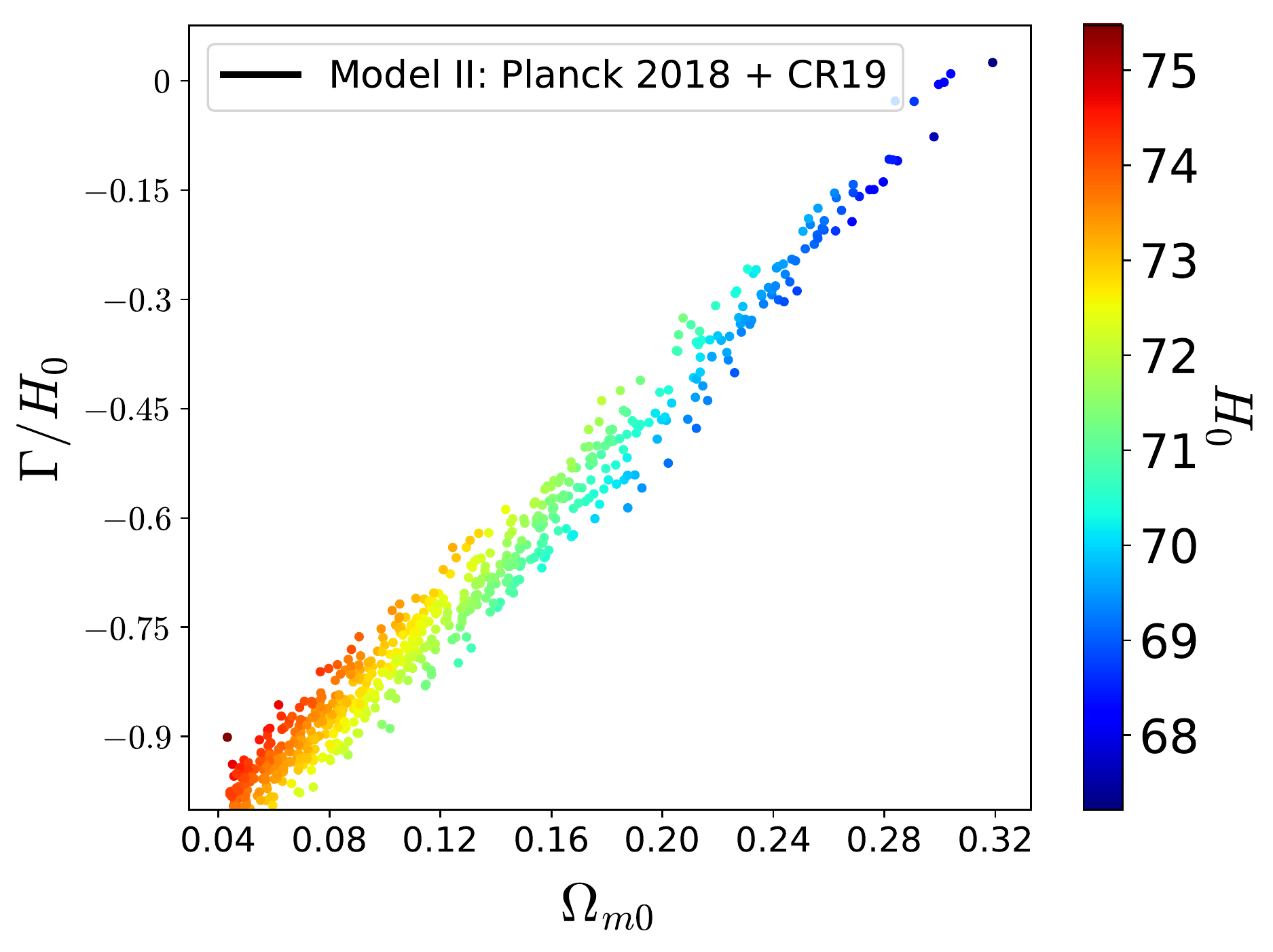}\\
\includegraphics[width=0.45\textwidth]{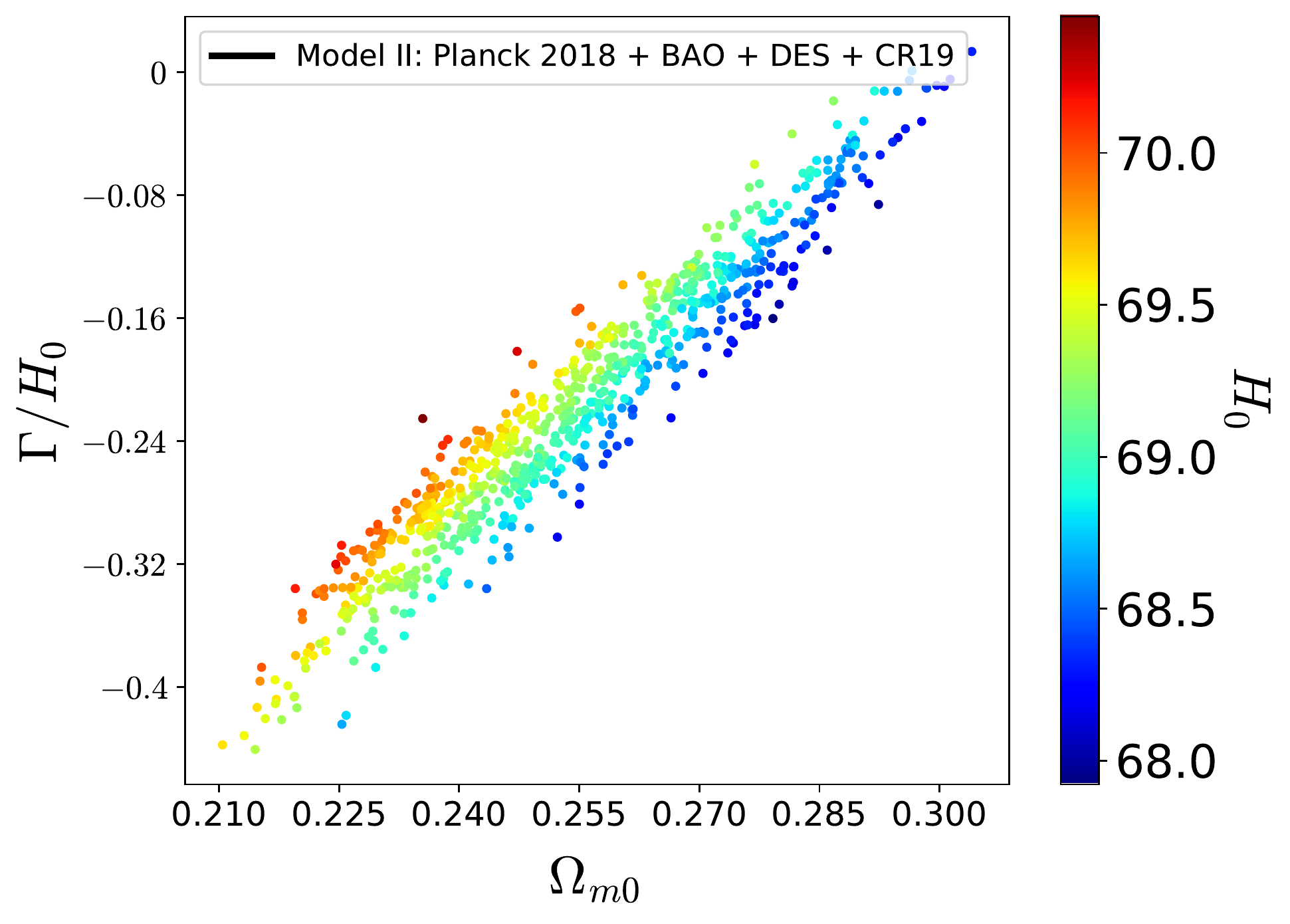}
\caption{3D scattered plots at 95\% CL in the plane $\Gamma/H_0$ vs $\Omega_{m0}$, coloured by the Hubble constant value $H_0$ for Model II. On the contrary of Model I, a strong positive correlation between $\Gamma/H_0$ and $\Omega_{m0}$, and a negative correlation between $\Gamma/H_0$ and $H_0$ are present. }
\label{fig-3d-model2}
\end{figure*}

\section{Results and analyses}
\label{sec-results}

In this section we present the observational constraints on the present metastable DE scenarios by considering data from Planck 2018 and other cosmological probes \ref{sec-data}.  Regarding  the initial conditions that are used during the analysis the situation is as follows. For the first model of our consideration, namely Model I, by following the notations of \cite{Doran:2003xq}, we have assumed adiabatic initial conditions. Now, although Model II represents a coupled cosmic scenario, if one assumes adiabatic initial conditions for the standard components, namely radiation and baryons, then the interacting dark fluids also follow the adiabatic initial conditions, see  \cite{Gavela:2009cy,Majerotto:2009np,Gavela:2010tm}.
The observational constraints for both the models are summarized in Tables \ref{tab:M1} (for Model 1) and Table \ref{tab:M2} (for Model 2). Further, the constraints on the $\Lambda$CDM cosmology (equivalently, $\Gamma  = 0$) have been shown in Table~\ref{tab:LCDM} for comparing the models with $\Gamma \neq 0$.  Additionally,
in  Figs. \ref{fig:model1} and \ref{fig:model2} we  present  the  corresponding  contour
plots (68\% and 95\% CL) for each model respectively.

\subsection{Model I}
\label{sec-results-m1}

Let us start with the presentation of the results for Model I. Using the data from Planck 2018 only (see second column of Table \ref{tab:M1}) we observe that the dimensionless parameter $\Gamma/H_0$ deviates from zero at more than $1\sigma$, and it is completely unconstrained at 95\% CL. We find that this parameter is correlated with most of the key parameters of the model. The fact that the $\Gamma/H_0$ is unconstrained from Planck 2018 data, can be easily verified if we look at Fig.~\ref{fig:CMB+matter_modelI}. We notice a strong positive correlation of the Hubble constant, $H_0$, with $\Gamma/H_0$, hence $H_0$ takes a relatively large value with very high error bars ($H_0 = 69.3_{- 3.5}^{+ 5.9}$, 68\% CL, Planck 2018) with respect to that of $\Lambda$CDM model (see Table \ref{tab:LCDM}). Therefore, in the context of Model I the $H_0$ measurement provided by Planck 2018 is compatible (within one standard deviation) with that of R19. Thanks to the geometrical degeneracy between $H_0$ and $\Omega_{m0}$ appeared in the CMB data, we also find that Model I prefers a lower value of the matter density. Indeed as we can see from Fig. \ref{fig-3d-model1},  there is a strong anti-correlation between $\Gamma/H_0$ and $\Omega_{m0}$.

Combining BAOs and Planck 2018 data we can place constraints on $\Gamma/H_0$ at 95\% CL, (see third column of \ref{tab:M1} and the 3D scattered plot of Fig.~\ref{fig-3d-model1}). This is due to the strong power of BAO data in constraining $\Omega_{m0}$ which anti-correlates with $\Gamma/H_0$. Notice, that in this case we have $\Gamma/H_0  = 0 $, i.e., in agreement with the $\Lambda$CDM model, within $1\sigma$. Further, regarding $H_{0}$ using Planck 2018+BAO dataset, we observe $2.6\sigma$
compatibility ($H_0 = 68.3_{-1.7}^{+ 1.6} $) with the corresponding value  obtained  R19, while in the case of the concordance $\Lambda$CDM model the difference is close to $\sim 4.4\sigma$.

Now let us test the combination Planck 2018+DES data.
The results of Planck 2018+DES combination are summarized in the fourth column of Table \ref{tab:M1}. In this case we have a lower limit of $\Gamma/H_0$, which is above zero (i.e. a cosmological constant model), at $2\sigma$ level, implying a decaying DE component. Concerning $\Omega_{m0}$, its best fit value becomes relatively low, namely $\Omega_{m0} = 0.263_{- 0.027}^{+  0.012}$ (68\% CL, Planck 2018+DES). Thanks to the three-parameter correlation shown in Fig. \ref{fig-3d-model1}, we find that
the best value of $H_0$ tends to that of R19 together, while the corresponding errors bars are quite large.

Now the  statistical results of the combined dataset Planck 2018+R19 are shown in the fifth column of Table \ref{tab:M1}. For this combination of data we find a strong indication of decaying DE with $\Gamma/H_0>0$ at more than $2\sigma$, namely we obtain $\Gamma/H_0 > 0.53$ at 95\% CL. These constraints are in very good agreement with those of Planck 2018+DES, showing a resolution of the tension with the cosmic shear data at the same time.

Finally, using Planck 2018+BAO+DES+R19 we present the corresponding results in the last column of Table \ref{tab:M1}. Also in this case $\Gamma/H_0$ deviates from zero at $2 \sigma$ and we observe $1\sigma$ compatibility of
all acquired parameter values with the corresponding values  obtained  from  Planck 2018+DES data.

Lastly, for a better understanding on the constraints on $H_0$ of different observational datasets, in Fig. \ref{fig:whisker} we present all of them in a whisker plot diagram, where  we display the constraints on $H_0$ from the observational datasets employed for this model as well as we show two different vertical bands referring to the constraints from Planck 2018 (the vertical grey band) \cite{Aghanim:2018eyx} and the local estimation (the vertical sky-blue band) from R19 \cite{Riess:2019cxk}.

\subsection{Model II}
\label{sec-results-m2}

The results of the observational constraints for the second model of our analysis; that is, for Model II, are shown in Table \ref{tab:M2} and in Fig. \ref{fig:model2}. In Fig. \ref{fig:model2}, for some of the key parameters of this model we show their one-dimensional posterior distributions and the 2-dimensional joint contours at 68\% and 95\% CL.

For Planck 2018 alone we find an indication of a $\Gamma/H_0$ different from zero at more than $1\sigma$. In fact, we have the upper limit $\Gamma/H_0 < -0.39$ at 68\% CL. This clearly shows that the transfer of energy from DM to DE is preferred by Planck 2018 data. However, at $2\sigma$, $\Gamma  = 0$ is back in agreement with the data. On the other hand, from Fig. \ref{fig:model2} we find a strong anti-correlation between $H_0$ and $\Gamma/H_0$, thus, as long as $\Gamma/H_0$ decreases, $H_0$ should increase. This fact is reflected by the Hubble constant constraint $H_ 0 = 70.3_{-  2.0}^{+  3.3}$ (68\% CL), which clearly shows that the tension on $H_0$ between Planck 2018 and R19 is solved within $2$ standard deviation.
Moreover, for this model, because of the flow of energy from DM to DE, we find a lower estimation of cold dark matter ($\Omega_{m0} = 0.18 ^{+0.07}_{-0.13}$ at 68\% CL) than its estimation within the $\Lambda$CDM model as obtained by Planck 2018 in~\cite{Aghanim:2018eyx}.
This is clearly expected for the geometrical degeneracy present in the CMB data: if we have less dark matter, we see a shift of the acoustic peaks and we need a larger $H_0$ value to have them back in the original position.

When BAO data are added to Planck 2018, thanks to the robust constraint BAO data give on the matter density $\Omega_{m0}$, we find that $\Omega_{m0}$ slightly increases with respect to the Planck 2018 alone case ($\Omega_{m0} =  0.242_{- 0.063}^{+    0.079} $ at 68\% CL), but it is still lower than the Planck 2018 value in the context of $\Lambda$CDM model \cite{Aghanim:2018eyx}. Due to the positive correlation between $\Omega_{m0}$ and $\Gamma/H_0$, as we can see from Figs.~\ref{fig-3d-model2} and~\ref{fig:model2}, we find that $\Gamma/H_0$ is in agreement with the zero value within one standard deviation. This means that $\Gamma/H_0$, i.e., the rate of energy transfer between the dark sectors, is in agreement with the expected value in the $\Lambda$CDM model. Hence, because of the very well known anti-correlation between $\Omega_{m0}$ and $H_0$, we see that the Hubble constant shifts towards lower value compared to its estimation from Planck 2018 alone, and moreover, its error bars are significantly decreased. Thus, the tension on $H_0$ slightly increases at $2.5\sigma$, but of course it is always less than the $4.4\sigma$ tension between Planck 2018 \cite{Aghanim:2018eyx} and the SH0ES collaboration \cite{Riess:2019cxk} within the $\Lambda$CDM scenario.  Moreover, because of the extraction method, the BAO data are not completely reliable in fitting extended DE models, as already pointed out in~ \cite{DiValentino:2019jae}.

We continue by considering the next two datasets Planck 2018+DES and Planck 2018+R19. For both cases since the tension between the datasets (Planck 2018, DES) and (Planck 2018, R19) is solved in this scenario, we can safely combine them, that means, we can consider the combined analysis Planck 2018+DES and Planck 2018+R19.
The results for Planck 2018+DES and Planck 2018+R19 are shown in the last two columns of Table \ref{tab:M2}.
For Planck 2018+DES we remark a really strong bound on $\Gamma/H_0$, which is lower than zero at more than $2\sigma$ and very well constrained. Since $\Gamma/H_0$ takes larger values than Planck 2018 and Planck 2018+BAO, and as we observe in Fig. \ref{fig-3d-model2} for the three parameter correlation, it follows a slightly larger value of $\Omega_{m0}$ and a smaller value of $H_0$ with respect to the previous cases. For this reason the Hubble constant tension with R19 is restored in this scenario at about $3.6\sigma$.
For Planck 2018+R19 we find a very strong upper limit on $\Gamma/H_0$, that is less than zero at several standard deviations. That means essentially we have an increasing DE scenario for this metastable DE model.
Concerning $\Omega_{m0} $ estimations, similarly to the previous cases, the matter density again decreases.

Finally, we combined all the datasets and showed the results in the last column of Table \ref{tab:M2}. Our results are similar to what we have observed with Planck 2018+DES. That means an indication of negative value of $\Gamma/H_0$ is supported by the combined data.

We refer to Fig. \ref{fig:whisker} showing the whisker plot of $H_0$ at 68\% CL with its measurements by different observational data. The whisker plot in Fig. \ref{fig:whisker} clearly shows how the tension on $H_0$ is alleviated for most of the data combination, with the exception of Planck 2018+DES.
In summary, within this metastable DE scenario, the energy density of DE is increasing, as reported by the observational data preferring a negative value for $\Gamma/H_0$.

\section{Summary and Concluding remarks}
\label{sec-discuss}

In this work we have investigated two metastable DE models by considering their evolution at the level of linear perturbations and constrain their parameter space in light of the latest observational data with a special focus on the CMB data from Planck 2018.
The consideration of perturbation equations is one of the main ingredients of our work and therefore the present article
generalizes earlier publications of \cite{Shafieloo:2016bpk,Li:2019san} where the perturbation equations of the metastable DE models were not considered.
Since the early time instability of any dark energy model can be visualized directly by investigating its equations at the perturbative level
implies that the inclusion of the perturbations equations are essential in understanding the actual dynamics of the DE model.
Additionally, there is a relation between the observational constraints of the explored models and the dynamical level, that means, whether the dynamics
of the model is considered at the background level or at background plus perturbative levels.
Concerning the observational data, we use the full
CMB measurements from final Planck 2018 release \cite{Aghanim:2018oex,Aghanim:2019ame}, BAO \cite{Beutler:2011hx,Ross:2014qpa,Alam:2016hwk}, DES \cite{Troxel:2017xyo, Abbott:2017wau, Krause:2017ekm} and a measurement of $H_0$ from SH0ES collaboration (R19) \cite{Riess:2019cxk}.
In order to investigate the present models, we have considered the following datasets and their combinations: Planck 2018 alone, Planck 2018+BAO, Planck 2018+DES and Planck 2018+R19. The inclusion of BAO to CMB is used to break the degeneracies between the parameters. For the last two cases, i.e., 2018+DES and Planck 2018+R19, the combination of Planck 2018 to either DES or R19 is possible since the tensions between these datasets are solved within these models.

For the first metastable DE model (\ref{model1}), we have summarized the results in Table \ref{tab:M1} and in Figs. \ref{fig:model1} and \ref{fig-3d-model1}. We remark that for all datasets we find $\Gamma/H_0 > 0$ which indicates that DE has a decaying nature within this context. While we mention that for Planck 2018 alone, $\Gamma/H_0$ remains positive at about 68\% CL, such evidence becomes stronger for the following combinations Planck 2018+DES and Planck 2018+R19. However, for Planck 2018+BAO, $\Gamma =0$ is consistent within 68\% CL. Additionally, we found that within this model, the tension on $H_0$ is mostly solved. Specifically, we notice that for Planck 2018 data alone, Planck 2018+DES and Planck 2018+R19, the tension on $H_0$ is significantly alleviated within $1\sigma$.  However, for Planck 2018+BAO, the tension on $H_0$ is just reduced at $2.6\sigma$ (see Fig. \ref{fig:whisker} for a better understanding).

The results of the second metastable DE model are shown in Table \ref{tab:M2} and Fig. \ref{fig:model2}. From the results, one can clearly conclude that, within this model scenario, $\Gamma/H_0 <0$ is preferred for all the data combination, with the exception of Planck 2018+BAO where $\Gamma =0$ is consistent within 68\% CL. So, for most of the observational data, an increasing of DE density (i.e., DM decays into DE) is favored. The tension on $H_0$ is alleviated for Planck 2018 within $2\sigma$. However, for Planck 2018+BAO it is weakened at $2.5\sigma$ and for Planck 2018+R19 it is completely solved.

Concerning the earlier publications of \cite{Shafieloo:2016bpk,Li:2019san}, the main improvements of the present work can be seen as follows.
First the inclusion of the perturbation equations of the metastable DE models generalizes the work of \cite{Shafieloo:2016bpk,Li:2019san} and second the
present work employs the CMB full likelihood analysis compared to those of \cite{Shafieloo:2016bpk,Li:2019san} where the CMB distance priors were used.
These differences naturally introduce some differences as far as the observational constraints are concerned,
specially on the estimation of the Hubble constant, $H_0$. We believe that our work offers a very transparent picture in alleviating
the so called Hubble constant tension. In fact, from Figs. \ref{fig:CMB+matter_modelI}, and \ref{fig:CMB+matter_modelII} one can understand how
the models behave on large scales. In particular, Fig. \ref{fig:CMB+matter_modelII} clearly demonstrates how the coupling parameter
plays an important role in order to quantify the behaviour of Model II on large scales.

Thus, based on the  observational data considered in this work and the results, specifically, focusing on the non-zero values of $\Gamma/H_0$ obtained from the presently used datasets, one may strongly argue that the metastable DE models should be investigated further with more data points, see for instance the updated data points in \cite{Park:2017xbl} as well as the upcoming observational datasets in order to arrive at a definite conclusion regarding their viabilities.
Moreover, as we have found that the metastable DE models with just an additional extra free parameter $\Gamma/H_0$ can solve quite efficiently the Hubble constant tension.

Last but not least, we would like to emphasize that the choice of the metastable DE models is not unique. Since the nature of DE is not purely understood, thus, there is no reason to exclude other metastable DE models beyond the present choices. For instance, some alternatives to the exponential choice of Model I can be considered. In a similar way, one could also generalize Model II by considering other functional forms. Although Model II describes an interacting scenario and similar choices are available in the literature; however, the exact functional form of the interaction rate is not yet revealed. Hence, we believe that metastable DE models should gain significant attention in the cosmological community due to the fact that within such models, the extrinsic properties of the universe do not come into the picture, only the intrinsic nature of DE plays the master role.

\section{Acknowledgments}
The authors are grateful to the referees for their comments and suggestions that improved the quality of the manuscript.
WY has been  supported by the  National Natural Science Foundation of China under Grants No. 11705079 and No. 11647153. EDV acknowledges support from the European Research Council in the form of a Consolidator Grant with number 681431. SP has been supported by the Mathematical Research Impact-Centric Support Scheme (MATRICS), File No. MTR/2018/000940, given by the Science and Engineering Research Board  (SERB), Govt. of India. SB acknowledges support from the Research Center for Astronomy of the Academy of Athens in the context of the program  ``{\it Tracing the Cosmic Acceleration}


\end{document}